\documentclass[
  aps,
  prx,
  preprint,
  reprint,
  superscriptaddress,
  amsmath,
  amssymb,
  nofootinbib
]{revtex4-2}

\usepackage[T1]{fontenc}
\usepackage{lmodern}
\usepackage{graphicx}
\usepackage{dcolumn}
\usepackage{bm}
\usepackage{microtype}
\usepackage{xcolor}
\usepackage{hyperref}
\usepackage{booktabs}
\usepackage{chngcntr}
\hypersetup{
  colorlinks=true,
  linkcolor=blue!60!black,
  citecolor=blue!60!black,
  urlcolor=blue!60!black
}

\newcommand{\beginsupplement}{%
  \setcounter{table}{0}
  \renewcommand{\thetable}{S\arabic{table}}
  \setcounter{figure}{0}
  \renewcommand{\thefigure}{S\arabic{figure}}
  \setcounter{equation}{0}
  \counterwithout{equation}{section}
  \renewcommand{\theequation}{S\arabic{equation}}
  \setcounter{page}{1}
  \renewcommand{\thepage}{S\arabic{page}}
}
\begin{document}
% ============================================================
% Main PRL manuscript
% ============================================================

\title{Probing the pairing symmetry of moir\'e graphene superconductors}

\author{Sayak Biswas}
\affiliation{Department of Physics, The Ohio State University, Columbus, Ohio 43210, USA}

\author{Rajdeep Sensarma}
\affiliation{Department
of Theoretical Physics, Tata Institute of Fundamental
Research, Mumbai 400005, India}

\author{Mohit Randeria}
\email{randeria.1@osu.edu}
\affiliation{Department of Physics, The Ohio State University, Columbus, Ohio 43210, USA}

%\date{\today}

\begin{abstract}
The pairing symmetry of magic-angle moir\'e graphene is a fundamental question that remains unresolved.
Combining experimental and theoretical inputs, we constrain the superconducting order parameters that can
emerge from the incommensurate Kekul\'e spiral (IKS) normal state on the hole-doped side of $\nu\!=\!-2$.
Imposing the additional experimental constraint of nodal superconductivity,
we are left with the task of distinguishing between singlet or triplet pairing, and of determining
the $\Vec{d}$-vector in the latter case.
We propose definitive tests to identify the pairing symmetry
based on two classes of experiments using the response of the superconducting state to Zeeman field orientation.
The first set of predictions is for spectroscopic and thermodynamic measurements
sensitive to low-energy excitations near the nodes.
The second set is for phase-sensitive measurements of topologically protected
Andreev bound states near boundaries, whose spectroscopy is shown to provide a ``smoking gun"
signature of the pairing symmetry.
\end{abstract}

\maketitle

%{\bf Introduction:}
\section{Introduction}
\label{sec:introduction}
Eight years after the discovery of superconductivity (SC) in
magic-angle twisted bilayer graphene (TBG)~\cite{Cao2018SC,Cao2018IN}, the pairing symmetry of its superconducting state remains unresolved.
SC has also been observed in magic-angle twisted trilayer graphene (TTG)~\cite{Hao2021,Park2021}, whose flat bands~\cite{Bistritzer2011} are essentially the same \cite{Khalaf2019} as those of TBG.
We collectively refer to these systems as moir\'e graphene ~\cite{Andrei2021,Balents2020,Nuckolls2024review} superconductors.

The pairing symmetry of a superconductor is a sharply posed question with a definite answer rigorously constrained by group theory and one which
can be pinned down unambiguously by experiment. By contrast, identifying the underlying ``mechanism" of SC, especially in strongly
correlated systems, is a more challenging task that may not a admit a decisive test.
In fact, knowledge of pairing symmetry can provide powerful clues to the mechanism.

Although a plethora of microscopic pairing mechanisms have been proposed for moir\'e graphene, comparatively little effort has been devoted
to systematically understanding its order parameter. Early analyses~\cite{Lake2022,Scheurer2020} of pairing symmetry were undertaken
before it was recognized that the normal state exhibits an
incommensurate Kekul\'e spiral (IKS)~\cite{kwan2021,wagner2022,nuckolls2023,kim2023}. The symmetries of this normal state
constrain the possible SC order parameters and play a central role in our analysis.

We use experimental and theoretical inputs to identify the symmetries of the nematic normal state with an IKS,
focusing on the hole-doped side of $\nu\!=\!-2$ where the most robust SC is seen in TBG and TTG.
Combined with the key experimental constraint of nodal superconductivity, we strongly limit the allowed order parameters.
The experimental inputs into our theory, discussed in detail below,
are derived from consistent results from a variety of probes
including transport, quantum oscillations, scanning tunneling microscopy (STM), and kinetic inductance measurements.

We first show that the key open questions related to the SC order parameter are: (1) Is it even parity, singlet or odd parity, triplet pairing?
Note that it is far from obvious that this is the relevant classification given the multiple bands with valley and sub-lattice degrees of freedom,
and higher dimensional irreducible representations of the point group symmetry of moiré graphene.
And (2) if it is a triplet superconductor, what is its $\Vec{d}$-vector? Here
$\hat{d}$ denotes the spin-space axis along which Cooper pairs have spin projection $S_z\!=\!0$~\cite{vollhardt1990}.

We next show that the best way to answer both questions is to exploit the dependence of experimental observables on the direction of
a Zeeman field. TTG is particularly well suited for these experiments, with an in-plane mirror symmetry that ensures that
magnetic fields that lie in the plane couple only to the electron spin and not to the flat band dispersion~\cite{Cao2021Nematicity,Park2022Robust}.

We make sharp testable predictions for two classes of experiments that can unambiguously distinguish
between all candidate pairing states and nail down the order parameter symmetry.
The first set of experiments relate to observables that are dominated by low energy excitations near the nodes in the energy gap.
This includes, amongst others, tunneling spectroscopy as a function of the in-plane orientation of the Zeeman field; see Fig.~1.

We propose a second set of experiments that are sensitive to the phase of the order parameter. These involve the creation of
topologically protected Andreev bound states (ABS)~\cite{Hu1994,Sauls2018} at a planar interface between the superconductor and
a (possibly gate-defined) insulator.
We predict that the ABS signature in the local density of states near the interface, which can be probed by STM, provides smoking gun
evidence to settle the two questions raised above; see Figs.~3 and 4.

Note that we focus on low-energy observables (with energy $\ll$ maximum gap or $T_c$) for which the predictions are robust, and
strong correlations (low density, narrow bandwidth) and fluctuation effects (short coherence length) do not affect the conclusions.
For a detailed summary of our results and their experimental implications, the reader is directed to the last section of the paper.

%\textbf{Normal state constraints:} {\color{blue} This section has been reorganized and strengthened.}
\section{Normal state constraints}
\label{sec:normal-state}
We begin with experimental and theoretical input on the normal state electronic structure and Fermi surface
of moir\'e graphene superconductors. Hartree Fock (HF) theory~\cite{kwan2021,wagner2022} predicted that a small
($\gtrsim 0.1 \%$) hetero-strain, ubiquitous in these materials, stabilizes an incommensurate Kekul\'e spiral (IKS) in the normal state,
a conclusion that has been tested by numerical techniques~\cite{wang2023,huang2026} beyond HF. The IKS state is characterized by
inter-valley coherence (IVC)~\cite{bultinck2020} that leads to a Kekul\'e pattern on the graphene scale, which in turn is modulated with
a wave-vector $|\bm{Q}|\!\sim\!1/a_m$ where $a_m$ is the moir\'e lattice spacing.

STM experiments on TBG~\cite{nuckolls2023} and TTG~\cite{kim2023} have imaged
the predicted IVC Kekule modulation on the graphene scale as well as the incommensurate spiral on the moir\'e scale
across the range of fillings of interest.
The IKS breaks all point group symmetries down to $C_2$ generated by two-fold rotation about $\hat{z}$.
This is consistent with the nematicity first seen in TBG~\cite{Cao2021Nematicity}, and recent
transport data~\cite{zhang2026} demonstrate $C_2$ symmetry in the normal state of TTG.

Although IKS breaks the symmetry under translations ${T}(\bm{R})$ by a moir\'e lattice vector $\bm{R}$, it nonetheless
preserves an effective translation~\cite{kwan2021}
$\widetilde{T}(\bm{R})\!=\!e^{i \bm{Q}\cdot\bm{R}\,\tau_z/2}\, {T}(\bm{R})$, where $\tau_z$ labels the valleys.
We thus have bands with crystal momentum $\bm{k}$
corresponding to the eigenvalue $e^{i\bm{k}\cdot\bm{R}}$ of $\widetilde{T}(\bm{R})$.

We next turn to the questions of the normal state spin symmetry
(since spin-orbit coupling (SOC) can be ignored in graphene) and time reversal.
HF predicts a spin-unpolarized IKS state for the $\nu\!=\!-2$ insulator~\cite{kwan2021,wagner2022,vafek2025}.
The IKS order survives upon hole doping this insulator, and inter-valley interactions stabilize~\cite{wang2025} a
spin-rotation invariant IKS ground state for a range of dopings $-3\!<\nu_c\!<\!\nu\!<\!-2$.
The singlet IKS state has a charge-Kekule distortion consistent with STM observations~\cite{nuckolls2023, kim2023},
while the other competitive state~\cite{wang2025} is a triplet IKS with a spin-Kekule modulation which would not be seen by STM.
Thus the normal state breaks $SU(2)_K\times SU(2)_{K'}$ down to a global $SU(2)_{\rm spin}$ and is time reversal (TR) invariant.

Finally, we discuss the normal state electronic structure upon hole doping the
$\nu\!=\!-2$ insulator. HF predicts a $2$-fold degenerate
Chern-zero band with a dispersion $\epsilon({\bf k})$ that leads to
a hole pocket centered around the $\Gamma$ point~\cite{wang2025}.
The low-energy spectral weight near $\Gamma$ is in marked contrast to the non-interacting
band structure~\cite{Bistritzer2011}, and is predicted at the HF
level~\cite{kwan2021,wagner2022,bernevig2022,vafek2025} and beyond~\cite{ledwith2025}.
The $2$-fold degeneracy of the bands arises from spin in these theories.
This is consistent with quantum oscillation experiments~\cite{Cao2018SC,Saito2021}
that observe $2$-fold degenerate Landau fans.
 %We will analyze below the SC order parameters that can emerge from the
 %spin rotation invariant IKS normal state. Its electronic structure on the
 %hole-doped side of $\nu\!=\!-2$ consists of a spin-degenerate hole pocket
 %around $\Gamma$ with zero Chern number. This is the simplest starting point
 %consistent with normal state experiments and theory.
%{\bf SC state constraints:}
\section{SC state constraints}
Experiments in the SC state lead to several important conclusions. First, ac Josephson
experiments~\cite{deVries2021} give direct evidence for a charge $2e$ (pair) condensate. Second, the nematicity observed in the
in-plane critical field and critical current in TBG~\cite{Cao2021Nematicity} and TTG~\cite{zhang2026,Cao2021Pauli, Park2022Robust} implies a $C_2$ symmetry. Third, there is no evidence for broken TR in the SC state.
%Specifically, a spin-polarized triplet SC state is inconsistent~\cite{Lake2022} with experiments.

There is strong evidence for point nodes in the SC gap, which will play a crucial role in our analysis.
The first evidence came from the $V$-shaped STM tunneling spectra in TBG~\cite{oh2021} and TTG~\cite{kim2022}.
Recent $c$-axis planar tunneling experiments in TTG \cite{park2026}, especially the $\sqrt{B}$-field dependence
(Volovik effect)~\cite{Volovik1993} of the density of states (DOS), gives strong evidence for nodes.
Linear in $T$ suppression of the superfluid stiffness in kinetic inductance measurements in TTG~\cite{banerjee2025}
reinforces the case for point nodes; similar measurements in TBG~\cite{tanaka2025} show a higher power law
that may be consistent with disorder effects on the nodes.
(Recent STM data on TTG~\cite{kim2026} show $U$-shaped spectra; we will comment on this below.)

To distinguish a node from a very small gap value, one needs an experiment that is sensitive to the phase of the order parameter.
The only such information in moir\'e graphene SCs comes from Andreev spectroscopy~\cite{oh2021,kim2022}.
Theoretical analysis~\cite{biswas2025} motivated by these data give strong evidence for Andreev bound states (ABS)
induced at the interface between the normal metal STM tip in contact with the sample.
These ABS exist only for a sign changing order parameter.

A striking feature of moir\'e graphene SCs is their large $T_c/E_F$ ratio or
$k_F\xi_0 \gtrsim 1$~\cite{Cao2018SC,tian2023},
reminiscent of the BCS-BEC crossover~\cite{randeria2014}
rather than the weak coupling limit $k_F\xi_0 \gg 1$.
The existence of gap nodes implies that these materials must be
on the BCS-side (``weak pairing regime") of the crossover, even though they are not in the extreme weak coupling limit.
On the BEC-side (``strong pairing regime") of the crossover, the energy gap is non-zero at all
$\bm{k}$ even if the order parameter has nodes~\cite{randeria1990}.

It is often argued that a violation of the Pauli (or Chandrashekhar-Clogston) limit the in-plane critical field \cite{Cao2021Pauli}
is a definitive signature against singlet SC. However, we caution that this expectation is based on the ratio of energy scales
computed in the extreme weak coupling limit, similar to $2\Delta/k_B T_c$. Quantitative violations of extreme weak coupling BCS results
could well arise (in part or whole) from strong correlation and fluctuation effects. We will thus focus in the following on proposing tests that are
{\it qualitatively} different for different symmetries, rather than comparing numbers that are hard to reliably compute in strongly correlated materials.

%\textbf{Symmetry constraints on order parameter:}
\section{Symmetry constraints on order parameter}
Armed with these constraints, we analyze the SC states that can arise out of a
spin-rotationally invariant IKS normal state. The order parameter matrix $\Delta(\bm{k})$ enters the pairing term through
    \begin{equation}
        \hat{H}_{\text{pair}} = \sum_{\bm{k}} \Delta_{\alpha\beta}\bigl(\bm{k}\bigr)c^\dagger_{\bm{k}\alpha}c^\dagger_{-\bm{k}\beta}   + \text{h.c.}
    \end{equation}
where $c^\dagger_{\bm{k}\alpha}$ creates an electron with crystal momentum $\bm{k}$ and spin $\alpha$; we
sum over spin indices.
We choose a gauge in which TR acts as
$\mathcal{T}c^\dagger_{\bm{k}\alpha}\mathcal{T}^{-1} = \bigl[i\sigma_y\bigr]_{\alpha\beta}c^\dagger_{-\bm{k}\beta}$ and
$\mathcal{C}_{2z} c^\dagger_{\bm{k}\alpha}\mathcal{C}_{2z}^{-1} = c^\dagger_{-\bm{k}\alpha}$.
Fermi statistics implies $\Delta^{\text{T}}\left(-\bm{k}\right)=-\Delta\left(\bm{k}\right)$, TR invariance leads to
$\left(i\sigma_y\right)\Delta^*\left(-\bm{k}\right)\left(i\sigma_y\right)^{\text{T}}=\Delta\left(\bm{k}\right)$, and
$\Delta\left(\bm{k}\right)\to \Delta\left(-\bm{k}\right)$ under $C_{2z}$.
Given the spin rotation symmetry of the normal state, the order parameter can be either
a singlet ($S\!=\!0$) or a triplet ($S\!=\!1$).

\textit{Spin singlet pairing}: Here $\Delta\left(\bm{k}\right) = \Phi_0(\bm{k})\bigl(i\sigma_y\bigr)$,
and Fermi statistics and TR imply that $\Phi_0(\bm{k})$ is a real function even in $\bm{k}$.
We write $\bm{k} = k\left(\cos\theta_{\bm{k}} , \sin\theta_{\bm{k}} \right)$, and
for small hole doping near $\nu = -2$ we can expand  $\Phi_0(\bm{k})$ about the
$\Gamma$ point in powers of $ka_m$. We find
        \begin{align}
            \Phi_0(\bm{k}) =& \Delta_0 + \Delta_2\left(ka_m\right)^2 \cos\left(2\theta_{\bm{k}} - 2\beta\right),
        \end{align}
where we ignore the higher, even angular momentum ($\ell$) contributions for small $ka_m$.
Now the $s$-wave amplitude must be small compared to the $d$-wave term
to obtain gap nodes. Even in this case, the nodes could be destroyed as $ka_m\!\to\!0$.
This may account for the change from nodal (V-shaped) to gapped (U-shaped) spectra~\cite{kim2022,kim2026}
with doping, as could the BCS-to-BEC crossover. However, given the preponderance of
experimental evidence for nodes, we focus in this paper only on the nodal, weak pairing regime.

\textit{Spin triplet pairing}: Using standard notation~\cite{vollhardt1990}
we write $\Delta\left(\bm{k}\right) = \Vec{d}\left(\bm{k}\right)\cdot\Vec{\sigma}\bigl(i\sigma_y\bigr)$. We use
bold symbols for 2D vectors in space and vector signs for 3D spin space. Fermi statistics and TR constrain
$\Vec{d}\left(\bm{k}\right)$ to be a real vector that is odd in $\bm{k}$.
Expanding $d_i\left(\bm{k}\right)$ $\left(i = x,y,z\right)$ near the $\Gamma$ point, and ignoring
higher odd $\ell$ terms, we find a $p$-wave gap function.

Symmetry considerations, together with the constraint that $\Vec{d}(\bm{k})$ must vanish at a node,
lead to the important conclusion that $\hat{d}$, the {\it direction} of the $\Vec{d}$-vector, must be
$\bm{k}$-independent; see Appendix~\ref{app:d_vector}.
We thus obtain
\begin{align}
            \Vec{d}\left(\bm{k}\right) = \Phi_1(\bm{k})\,\hat{d}; \ \ \ \ \Phi_1(\bm{k}) = \Delta_1\,ka_m \cos\left(\theta_{\bm{k}} - \beta\right).
\label{Eq:d-vector}
\end{align}

This raises the question of what determines the direction of the $\Vec{d}$-vector.
The only interaction that can pin $\hat{d}$  is a very small SOC, an energy scale ignored in our
analysis. It is well known from Leggett's analysis~\cite{vollhardt1990} of superfluid $^3$He that
the effects of a tiny SOC are amplified by the macroscopic condensate. In Appendix~\ref{app:d_vector}, we discuss
symmetry constraints on $\hat{d}$ in the presence of a tiny SOC that preserves mirror $M_z$ symmetry.
However, without committing to microscopic details of very small energy scales,
we will consider an arbitrary $\Vec{d}$-vector orientation and show how it can be determined experimentally.

{\it Spin structure of the order parameter:}
We now turn to the key questions of singlet versus triplet pairing and
the experimental determination of the $\Vec{d}$-vector in the latter case.
We address these issues by examining the response of the system to small
Zeeman fields focussing on sharp signatures as a function of its direction.

The mirror symmetry ${\cal M}_z$ in TTG protects the electronic structure of the flat bands,
and an in-plane $\Vec{B}$ couples only to the spins. We shall only discuss the effects of small Zeeman fields,
with $B\ll$ the maximum gap or $T_c$, where we set $g\mu_B = 1$ and ignore $(\Vec{d}\cdot\vec{B})^2$ terms.
We thus restrict ourselves to small fields that probe the order parameter but do not affect it.

\section{Effective low-energy Hamiltonian}
\label{sec:low_energy_ham}
%{\bf Effective low-energy Hamiltonian:}
We use four component spinors $(\Psi_{\mathbf{k}}, \Psi^*_{-\mathbf{k}})^T$
with $\Psi_{\mathbf{k}}=(c_{\mathbf{k}\uparrow},c_{\mathbf{k}\downarrow})^T$.
We make unitary transformations
(i) ($\Psi_{\mathbf{k}}, (i\sigma_y)\Psi^*_{-\mathbf{k}})^T$ for a singlet SC, and to
(ii) $(\Psi_{\mathbf{k}}, {\hat{d}}\cdot\Vec{\sigma}\,(i\sigma_y)\Psi^*_{-\mathbf{k}})^T$ in the triplet case.
The $4\!\times\!4$ Bogoliubov-deGennes (BdG) Hamiltonian then reduces to the same form in both cases
\begin{equation}
            \mathcal{H}({\bm{k}}) = {\bm u}({\bm{k}})\cdot{\bm \tau}\otimes \sigma_0, \ \ \ {\bm u}({\bm{k}})= \left(\Phi_S({\bm{k}}), 0, \xi({\bm{k}})\right),
\label{Eq:bdg1}
\end{equation}
where $\xi({\bm{k}})\!=\!\epsilon({\bm{k}})\!-\!\mu$ is the electronic dispersion
$\epsilon({\bm{k}})$ measured from the chemical potential $\mu$,
the subscript $S\!=\!0$ ($S\!=\!1$) denotes the singlet (triplet) order parameter, and $\sigma_0 = \mathbb{I}_{2\times 2}$.
TR implies a real $\Phi_S({\bm{k}})$ leading to a chiral symmetry $\tau_y$, whose implications we discuss below.

We analyze the response to a Zeeman field $\Vec{B}$ using
the Green's function $G_0(\bm{k}, z) = (z-\mathcal{H}_B\left(\bm{k}\right))^{-1}$, where
$\mathcal{H}_B = \mathcal{H} + \mathcal{H}^\prime$. Here $\mathcal{H}$ is defined in \eqref{Eq:bdg1} and
\begin{eqnarray}
\mathcal{H}^\prime = \left\{ \begin{array}{rcl}
 \tau_0\, \otimes\, {\Vec{B}}\cdot{\Vec{\sigma}} \quad \quad \quad \quad \quad \quad \, \ \ & \mbox{for} & S\!=\!0
\\
 \tau_0\, \otimes\, {\Vec{B}_{\parallel}}\cdot{\Vec{\sigma}}  + \tau_z\, \otimes\, \Vec{B}_{\perp}\cdot\Vec{\sigma} & \mbox{for} & S\!=\!1
\end{array}\right.
\label{Eq:Zeeman}
\end{eqnarray}
Here $\tau_0 = \mathbb{I}_{2\times 2}$, and $\Vec{B}_{\parallel} = ( {\Vec{B}}\cdot\hat{d} )\,\hat{d}$ and
${\Vec{B}_{\perp}} = \Vec{B} - \Vec{B}_{\parallel}$ are
components of $\Vec{B}$ parallel/perpendicular to $\hat{d}$.
%{\bf Bogoliubov quasiparticle spectrum:}

We look at the poles of $G_0$ to study the impact of the Zeeman field
on the spectrum of Bogoliubov excitations.
For the singlet SC the spectrum is given by $E_{\bm{k}} = \sqrt{\xi^2({\bm{k}})+\Phi_0^2({\bm{k}})} + \sigma B$ for $\sigma = \pm1$,
independent of the in-plane orientation of $\Vec{B}$.

For a triplet SC, on the other hand, the excitation spectrum depends in general on the direction of $\Vec{B}$
relative to the $\Vec{d}$-vector, and is given by
\begin{equation}
E^2_{\bm{k}} = \xi^2({\bm{k}})+\Phi_1^2({\bm{k}})+ B^2 \pm 2\sqrt{B^2\xi^2({\bm{k}}) + B_\parallel^2\Phi_1^2({\bm{k}})}
\label{Eq:spectrum}
\end{equation}
where $\Vec{B}_{\parallel,\perp}$ were defined below eq.~\eqref{Eq:Zeeman}.

The temperature and energy dependence of various experimental observables depends in a crucial way on the excitation spectrum of the system.
Thus even before we present any detailed calculations we can reach general conclusions from the results above.
{\it If an experiment finds results that vary with the orientation of an in-plane Zeeman field, it rules out both a pure singlet SC
order parameter and a triplet SC with a $\Vec{d}$-vector parallel to $\hat{z}$}.
We will propose sharp tests below to distinguish between singlet and triplet and show how one can pin down
the $\Vec{d}$-vector experimentally.

%{\bf Low energy density of states:}
\section{Low energy density of states}
A direct probe of the excitation spectrum is their DOS
$N(E) = - {\rm Im} \sum_{\bm{k}} G_0({\bm k}; z\!=\!E+i0^+)/\pi$ probed by tunneling.
In the absence of a field, the low-energy DOS is given by $N(E) \approx N_0\left|E\right|$,
where $N_0 \sim 1/(v_F v_{\Delta})$. This is obtained by noting that
the low energy DOS is dominated by excitations near the zero-field node, where
$\xi(\bm{k}) \approx v_F\delta k_\perp$, with $v_F$ the Fermi velocity,
and $\Phi_S(\bm{k}) \approx v_\Delta \delta k_\parallel$, with $v_\Delta$ the ``gap velocity".
Here $\delta k_\perp$  and $\delta k_\parallel$ are the deviations from the node perpendicular and parallel to the Fermi surface (FS).
(Here we assume, for simplicity, that the nodal direction in $\Phi_S(\bm{k})$ is perpendicular to the FS; the
 final results are independent of this assumption. See Supplementary Info).

For a singlet SC in a Zeeman field $\Vec{B}$, the DOS changes to $N(E) = N_0\left( \left| E+B \right| + \left| E - B \right| \right)/2$,
leading to $N(0) = N_0 B$~\cite{yang1998}. As emphasized earlier, these results are independent
of the in-plane direction of $\Vec{B}$.

\begin{figure}[t]
    \centering
    \includegraphics[width=\columnwidth]{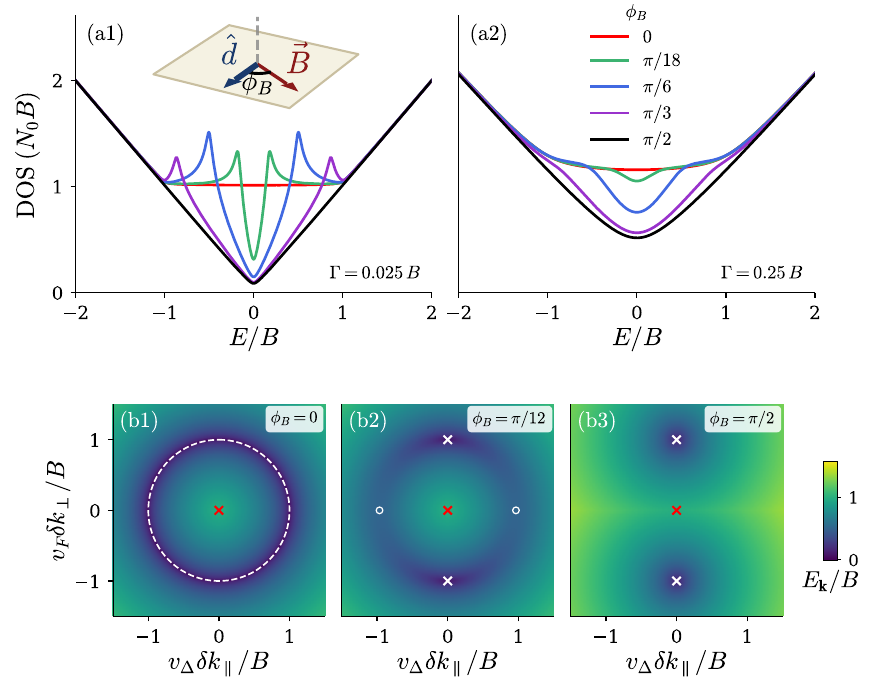}
    \caption{\textbf{Low-energy DOS in a Zeeman field for triplet pairing.} (a1) Variation of the DOS $N(E)$
    with the angle $\phi_B$ between $\Vec{B}$ and $\hat{d}$.  $N(E)\!\sim\!|E|$ is nodal for
    $\phi_B=\pi/2$, whereas it is finite with $N(0)\!\sim\!B$ when $\phi_B=0$.
     For $0<\phi_B<\pi/2$, the DOS remains nodal and exhibits low-energy van Hove singularities at $E=\pm B\sin\phi_B$; see text.
     (a2) These singularities are wiped out by a broadening $\Gamma$, but $N(0)$ shows a variation with $\phi_B$.
     (b1--b3) Evolution of the nodes in the Bogoliubov spectrum~\eqref{Eq:spectrum}. The zero-field node is at $(0,0)$ (red cross).
     For $\phi_B=0$, the nodes are a (white dashed) circle of radius $B$ in the ($v_\Delta\delta k_\parallel,v_F\delta k_\perp$)-plane;
     see text. For $\phi_B\neq0$, there are only two point nodes (white crosses). For $0<\phi_B<\pi/2$, the van Hove singularities
     are indicated by the white open symbols.
    }
    \label{fig:1}
\end{figure}

For the triplet case, the results depend on the in-plane direction of $\Vec{B}$ relative to the $\Vec{d}$-vector,
characterized by angle $\phi_B$. Two limiting cases lead to particularly simple results.
(i) When $\Vec{B} \parallel \hat{d}$ ($\phi_B = 0$), we obtain the same result as the singlet case, namely
$N(E) = N_0\left( \left|E+B\right|+ \left|E-B\right| \right)/2$, with $N(E) = N_0 B$.
(ii) For $\Vec{B} \perp \hat{d}$ ($\phi_B = \pi/2$), we find that $N(E) = N_0\left|E\right|$
remains unchanged from its zero field form.

For a triplet SC with its $\Vec{d}$-vector perpendicular to the plane ($\hat{d} = \hat{z}$), case (ii) applies for all in-plane orientations of the
Zeeman field. In this case the DOS of the triplet SC is unaffected by $\Vec{B}$.

On the other hand, when the $\Vec{d}$-vector lies in the plane the DOS
shows non-trivial dependence on the in-plane field orientation; see Fig.~1(a1).
Note that we plot the DOS on a very small energy scale $B \ll \Delta$ the maximum gap.
While the $\phi_B = 0$ and $\phi_B = \pi/2$ results are
just as expected from the two cases discussed above, the evolution at intermediate $\phi_B$ shows a new low-energy scale at
which ``coherence peaks" appear. To understand this, we analyze the evolution of the nodes with changing
$\phi_{B}$; see Fig.~1(b1,b2, b3).
Precisely at $\phi_B = 0$, each zero-field point node expands into a field-induced elliptical
``Bogoliubov Fermi contour" with major/minor axes that scale with $B$, which appears as a circle in the
scaled variables used in Fig.~1(b1).
This gives rise to the finite DOS $N(0)\!\sim\!B$.
For any $\phi_B \neq 0$, however, there are only two nodal points [Fig.~1(b2,b3)] and
a small gap opens up everywhere else. The van Hove singularity associated with this gap leads to
the low energy scale $E= B\sin\phi_B$ seen in Fig.~1(a1); for details see Appendix~\ref{app:DOS}.

In real materials, spectral features are invariably broadened by disorder and inelastic scattering.
In tunneling spectroscopy this is often modeled by a phenomenological broadening parameter $\Gamma$.
In Fig.~1(a2), we show how $\Gamma$ may wipe out the coherence peaks, but we still see a characteristic filling of
the low energy DOS as a function of $\phi_B$.
%{\bf Spin susceptibility:}
\section{Spin susceptibility}
The linear response to a Zeeman field can be quantified by the spin susceptibility tensor $\chi_{ij}$.
For the singlet SC, the response is {\it isotropic} with $\chi_{ij} = \chi(T) \delta_{ij}$. Quite generally, for a nodal SC, $\chi(T) \sim T$
as $T\rightarrow 0$. This power law is robust on the entire BCS-side of the crossover (weak pairing regime), independent of $k_F\xi_0$,
and also independent of interaction corrections, which only affect pre-factors. In the weak coupling BCS limit one can easily obtain an explicit formula for the full $T$-dependence of $\chi(T)$ in terms of the Yosida function $Y(T)$ with a nodal gap~\cite{vollhardt1990}.

The triplet SC has an {\it anisotropic} spin susceptibility
\begin{equation}
        \chi_{ij} = \chi_{\parallel}(T) \hat{d}_i\hat{d}_j  + \chi_\perp(T)  \bigl(\delta_{ij} - \hat{d}_i\hat{d}_j\bigr)
        \label{Eq:triplet-chi}
\end{equation}
written in terms of projection operators parallel and perpendicular to $\hat{d}$.
On the BCS-side of the crossover (weak pairing regime) $\chi_{\parallel}\sim T$ vanishes while $\chi_\perp$ remains finite as $T\rightarrow 0$.
(In the extreme BCS limit one can write explicit formulae $\chi_{\parallel}$ in terms of $Y(T)$ and $\chi_{\perp}$ in terms of
the normal state DOS).

In bulk quantum materials the spin susceptibility is probed by the NMR Knight shift, and it has been measured by optical experiments in some TMDs.
However, it appears that these experiments may not be feasible for moiré graphene~\cite{Shan}.

%{\bf Topologically protected Andreev bound states:}
\section{Topologically protected Andreev bound states}
We now turn to Andreev bound states (ABS) at an interface between a nodal superconductor and an insulator.
Before analyzing the problem with an interface, we first focus on
the {\it bulk} topological invariant which gives insight into
the conditions under which we obtain topologically protected ABS at $E=0$.
This has two important consequences.
(a) The existence of the ABS is robust on the entire BCS-side (weak pairing regime) of the crossover, and not just in the extreme weak coupling BCS limit.
(b) The results do not depend on the detailed nature of the boundary conditions (b.c.'s) for smooth interfaces.

To understand the bulk topological invariant associated with the Hamiltonian~\eqref{Eq:bdg1}, we
fix $k_y = k_y^0$, so that $\mathcal{H}{(k_x,k_y^0)}$ describes an effective 1D problem in the $x$-direction.
For almost all $k_y^0$ (except for a set of measure zero that go through a point node) the 1D system is gapped.
Writing ${\bm u} = u\, \widehat{{\bm u}}$ in eq.~\eqref{Eq:bdg1},
we look at the winding number ${\cal W}$
of the unit vector $\widehat{{\bm u}}{(k_x,k_y^0)}$ as $k_x$ varies across the Brillouin zone
to see if the 1D system (for fixed $k_y^0$) is topologically non-trivial; see Fig.~2.

For a $k_x$-path that does not cross the Fermi surface ($\xi({\bm k})=0$), the $z$-component of the $\widehat{{\bm u}}$ has a fixed sign,
and thus $\widehat{{\bm u}}$ does not wind around the unit circle in the $(x,z)$-plane; lower path in Fig.~2. On the other hand, for a $k_x$-path that crosses the Fermi surface,
one can obtain non-trivial winding. When the $k_x$-path crosses the Fermi surface twice, the
${\cal W} = 1$ when $\Phi_S({\bm{k}})$ has opposite signs at two Fermi points (top path in Fig.~2), but vanishes when
the order parameter has the same sign at the two Fermi crossings (middle path in Fig.~2).

 \begin{figure}[t]
    \centering
    \includegraphics[width=\columnwidth]{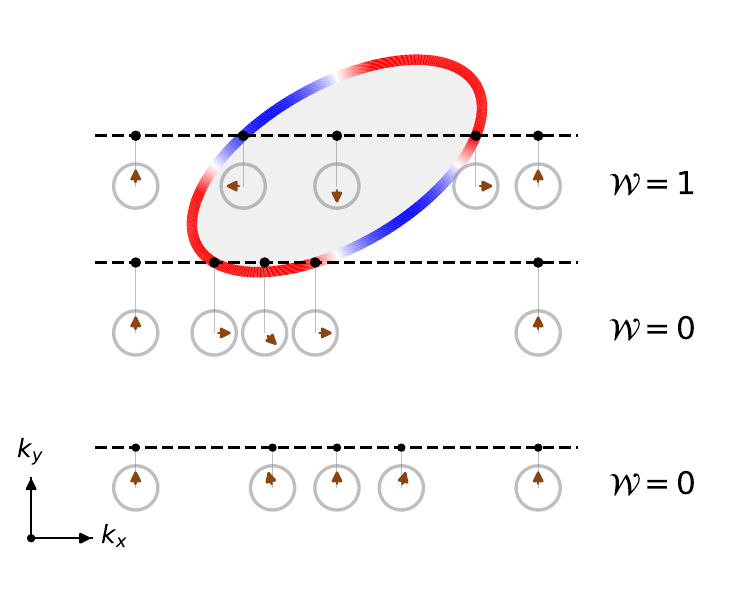}
    \caption{\textbf{Bulk topological invariant for ABS:} Winding of the unit vector $\widehat{{\bm u}}$ in the Hamilotonian~\eqref{Eq:bdg1}
    as $k_x$ traverses the Brillouin zone for three fixed values of $k_y$. Positive/negative signs of the SC order parameter are shown as red/blue on the
    Fermi surface (FS). The FS crossings on the top path have opposite sign order parameters leading to a winding number ${\cal W}=1$. The other two paths
    have ${\cal W}=0$. A non-zero winding in the bulk implies ABS at the boundary.}
    \label{fig:2}
\end{figure}

The sign change condition thus determines if we have a gapless boundary mode, the ABS, at the interface between the SC and a metal or insulator,
as first noted from BdG solutions in the weak coupling BCS limit~\cite{Hu1994,kashiwaya2000}
The topological argument clearly shows that $E=0$ ABS exist for all $\Delta/\mu$ or $k_F\xi_0$
in the entire BCS-side of the crossover (weak pairing regime). In the BEC-like (strong pairing) regime, however, the chemical potential lies outside the band
and thus $\xi({\bm{k}})\!=\!\epsilon({\bm{k}})\!-\!\mu \neq 0$ for all ${\bf k}$.

Although the 2D BdG Hamiltonian has both TR and particle-hole (PH) symmetries, these no longer survive for a fixed $k_y^0$.
 The combination of TR and PH is a chiral symmetry that survives in
the 1D problem, which is thus in the AIII class with a $\mathbb{Z}$ topological invariant \cite{sato2011,Schnyder2011}.
%While a smooth interface with a conserved $k_y$ underlies their topological protection at $E=0$, we emphasize that
%the ABS persist even in the presence of interfacial disorder~\cite{matsumoto1995,Fogrlstrom1997,Kalenkov2004}, but
%are shifted away from $E=0$ {\color{blue} and may acquire a width.}

 \begin{figure}[t]
    \centering
    \includegraphics[width=\columnwidth]{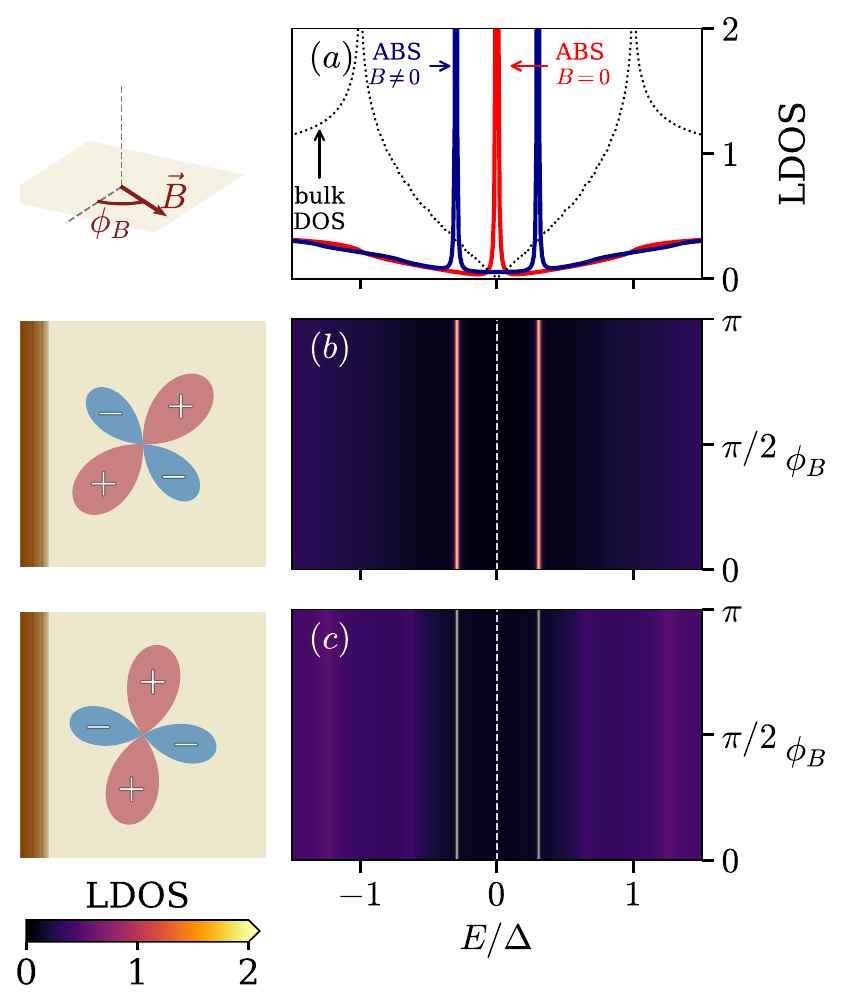}
    \caption{\textbf{Isotropic Zeeman splitting of ABS for singlet pairing.} (a) LDOS near the boundary shows a $E\!=\!0$
    peak arising from Andreev bound states (ABS) which splits into two peaks at $E=\pm B$ in a field. These LDOS are
    contrasted with the bulk DOS.
    (b,c) The ABS splitting is independent of the direction $\phi_B$ of the in-plane field $\Vec{B}$ as seen from the
    false-color plots of the LDOS as a function of $E$ and $\phi_B$ for two orientations of the $d$-wave order parameter with respect to the interface: (b) $\Delta \cos\left(2\theta_{\bm{k}}-\pi/2\right)$ and (c) $\Delta \cos\left(2\theta_{\bm{k}}-\pi/8\right)$.
    $B = 0.3 \Delta$ and all LDOS are calculated at a small distance $0.2\pi/k_F$ from the interface. The nematic nodal SC must have
    in addition to the $d$-wave a small $s$-wave amplitude [see text below eq.~(2)], which we ignore for our numerics as it does not impact the results qualitatively.
}
    \label{fig:3}
\end{figure}

\begin{figure*}[t]
    \centering
    \includegraphics[width=\textwidth]{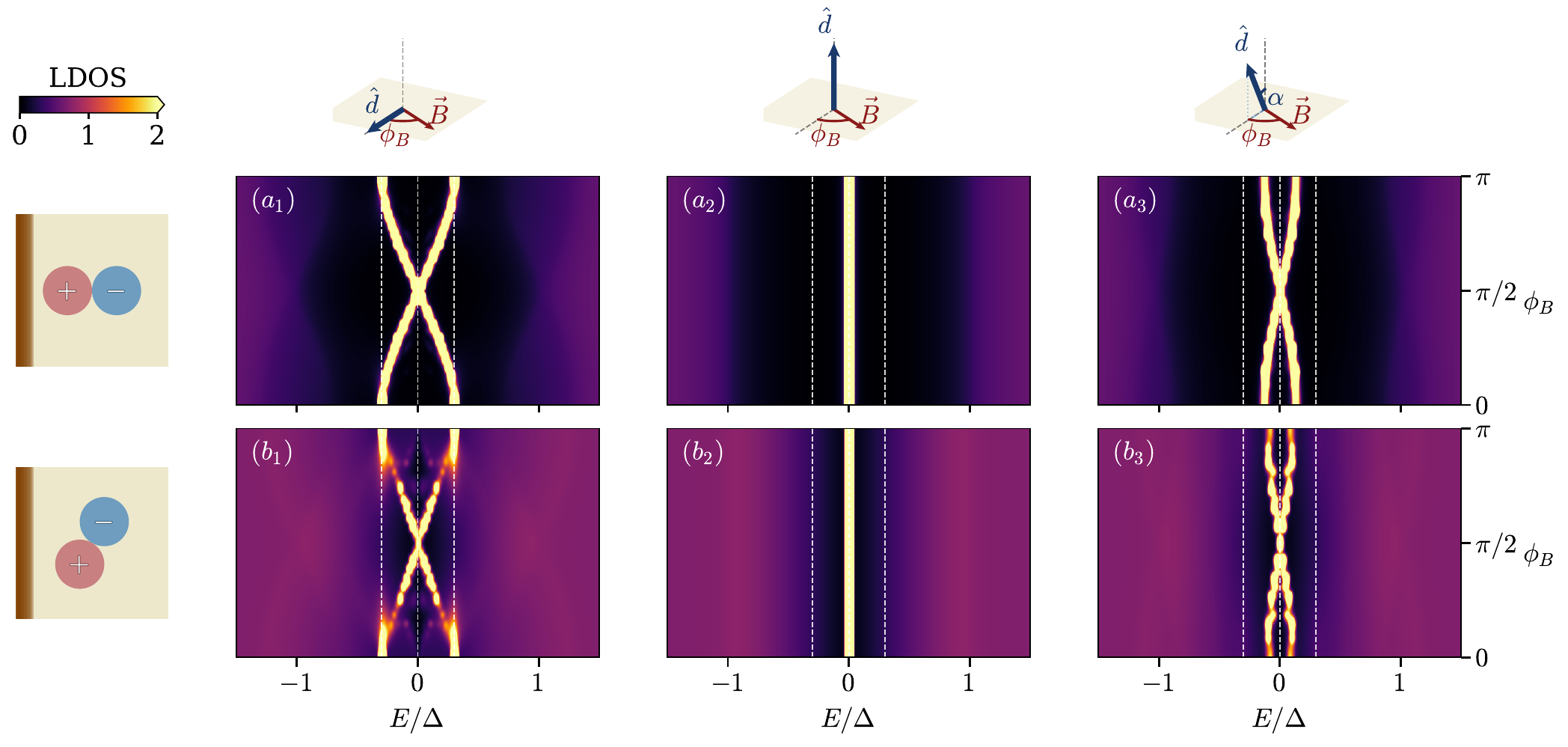}
    \caption{\textbf{Anisotropic Zeeman splitting of ABS for triplet pairing.} The LDOS near the boundary plotted as a function of
    $E$ and the in-plane field $\Vec{B}$ direction $\phi_B$ for three orientations of $\hat{d}$. (1) In-plane $\hat{d}$,
    (2) out-of-plane $\hat{d}$, and (3) $\hat{d}$ at an angle $\alpha= \pi/6$ with respect to the normal to the plane.
    Results are shown for two different orientations of the $p$-wave order parameter:
    (a) $\Delta\cos\left(\theta_{\bm{k}}\right)$ and (b) $\Delta\cos\left(\theta_{\bm{k}}-\pi/3\right)$.
    In (a1,b1) the ABS splits into two peaks at $E=\pm B$ when $\Vec{B}\parallel\hat{d}$ ($\phi_B=0$), while there is no splitting when $\Vec{B}\perp\hat{d}$ ($\phi_B=\pi/2$). In (a2,b2), $\hat{d}$ is always perpendicular to $\Vec{B}$, and hence there is no splitting.
    In (a3,b3), the magnitude of the splitting lies between those in cases (1) and (2). $B=0.3\Delta$ and all LDOS are calculated at at a small distance $0.3\pi/k_F$ from the interface; see text.}
    \label{fig:4}
\end{figure*}
%{\bf Effect of Zeeman field on ABS spectroscopy:}
\section{Effect of Zeeman field on ABS spectroscopy}
We use a Zeeman field $\Vec{B}$ to distinguish between a singlet and triplet SC, and to probe the $\Vec{d}$-vector in the latter case.
 Toward this end, we compute the local tunneling DOS near the interface where the ABS exist
and look at the response of the ABS spectrum to $\Vec{B}$.

The topological stability of the ABS implies that details of the boundary conditions (b.c.'s) do not matter
so long as we have a smooth interface between SC $(x > 0)$ and, say, vacuum $(x < 0)$ with $k_y$ conserved.
We solve for the {\it half-space Green's function} (GF) $G(x,x^\prime)$ with a
hard wall b.c.~$G(x,0) = G(0,x^\prime) = 0$.
For simplicity we suppress all GF arguments that are not necessary for our purposes;
here we omit $k_y$ and $z$.
We find the result
\begin{equation}
G(x,x^\prime) = G_0(x-x^\prime) - G_0(x)G_0(0)^{-1}G_0(-x^\prime)
\label{Eq:half-space-GF}
\end{equation}
which obeys the GF equation as well as the b.c.; see also ref.~\cite{matsumoto1995,Suman2026}.
Here $G_0(x)$ is the Fourier transform
of the {\it translationally invariant bulk GF} defined above eq.~\eqref{Eq:Zeeman}.

{\it Singlet Pairing:}
The form of the Zeeman coupling in eq.~\eqref{Eq:Zeeman} allows us to express the bulk GF $G_0$ for the $S=0$ SC as a sum of two terms
describe excitations with spins aligned parallel and antiparallel to $\Vec{B}$ with energies shifted by $\pm B$.
Using eq.~\eqref{Eq:half-space-GF}, we find that the half-space Green's $G$ function also has the same structure. We write this using condensed notation as
\begin{equation}
        G(z) = G_{B=0}(z-B)\otimes \mathbb{P}_{+} + G_{B=0}(z+B)\otimes \mathbb{P}_{-}
   \label{Eq:GF-singlet}
\end{equation}
where we omit all arguments except the complex energy $z$.
Here $\mathbb{P}_{\pm}\!=\!(\mathbb{I} \pm \widehat{B}\cdot\Vec{\sigma})/2$
are projection operators for spins parallel/antiparallel to $\Vec{B}$, and $G_{B=0}(z)$ are
zero-field $2\times 2$ GFs in $\tau$-space.

It is clear from eq.~\eqref{Eq:GF-singlet} that all excitations, including the $E=0$ ABS poles of the
half-space GF near the boundary, are Zeeman split by
$\pm B$, and the spectrum is independent of the in-plane direction of $\Vec{B}$.
To see these effects clearly, we show in Fig.~3 the local density of states (LDOS)
$N({\bm r},E) = - {\rm Im} G({\bm r},{\bm r}^\prime\!=\!{\bm r}; z\!=\!E+i0^+)/\pi$.
We probe the ABS using LDOS at a short (fraction of $k_F^{-1}$) but finite distance away from the boundary
as their wave function vanishes at the hard wall.
Far ($\gg \xi_0$) from the boundary, the LDOS is essentially the same as the bulk DOS.

In Fig.~3(a) we plot the LDOS for ${\bm r}$ near the interface, where we clearly see the zero-field ABS peak at $E=0$
that splits into $E=\pm B$ peaks in a Zeeman field. Also shown for comparison is the bulk DOS, identical to the  LDOS far from
the interface. From Figs.~3(b,c) we see that the Zeeman splitting of the ABS is independent of the angle $\phi_B$ between the field and the
interface normal. The only difference between the two panels is that in (b) the $d$-wave order parameter is oriented at $\beta\!=\!\pi/4$ to the
normal, while in (c) it is at an angle $\beta\!=\!\pi/16$. The spectral intensity of the ABS diminishes as $\beta$ decreases from $\pi/4$
and eventually vanishes at $\beta\!=\!0$, an orientation at which no ABS are supported~\cite{Hu1994,kashiwaya2000}.

{\it Triplet Pairing:}
In the general case, when $\Vec{B}$ makes an arbitrary angle with $\hat{{d}}$, the Zeeman coupling of eq.~\eqref{Eq:Zeeman} does not permit
a simple analytical result. We then use the procedure outlined in Appendix~\ref{app:ABS} to compute the half-space GF and deduce from it the ABS spectrum
in a Zeeman field. There are, however, two special orientations of $\Vec{B}$ with respect to $\hat{d}$ for which we can gain
analytical insight.

Case 1: When $\Vec{B} \parallel \hat{{d}}$, we have $\Vec{B}_{\parallel}\!=\!\Vec{B}$ and $\Vec{B}_{\perp}\!=\!0$ so
that the triplet $\mathcal{H}^\prime$ of eq.~\eqref{Eq:Zeeman}
is identical to the singlet case. Thus $G(z)$ is again given by eq.~\eqref{Eq:GF-singlet}, and the entire spectrum including
ABS are split by $\pm B$ when $\Vec{B} \parallel \hat{{d}}$.

Case 2: For $\Vec{B} \perp \hat{{d}}$, we have $\Vec{B}_{\parallel} = 0$ and $\Vec{B}_{\perp} = \Vec{B}$,
and the $\tau_z$ term in $\mathcal{H}^\prime$ comes into play, effectively shifting the
chemical potential $\mu$ by $\pm B$. We then find
\begin{equation}
G(z) = G_{B=0}(z;\mu-B) \otimes \mathbb{P}_{+} + G_{B=0}(z;\mu+B) \otimes \mathbb{P}_{-}
 \label{Eq:GF-B-perp-d}
\end{equation}
where $G_{B=0}$'s are zero-field $2\times 2$ GFs in $\tau$-space with
all arguments suppressed except for the complex energy $z$.
We see from eq.~\eqref{Eq:GF-B-perp-d} that the spectrum, and in particular the ABS, are not split by the Zeeman field
when $\Vec{B} \perp \hat{{d}}$.

Armed with these limiting cases we look at the results presented in Fig.~4 where the rows correspond to two different
orientations of the $p$-wave order parameter and the columns to possible orientations of $\hat{d}$. As in the singlet case, the
order parameter orientation only affects the intensities, but not the important spectral features of the ABS.
Consider first an in-plane $\Vec{d}$-vector (left column). We see from Fig.~4(a1) that the ABS energies are highly sensitive to the
angle $\phi_B$, varying from a maximal splitting of $\pm B$ when $\phi_B\!=\!0$ to a double degeneracy at $\phi_B\!=\!\pi/2$,
consistent with the two cases discussed above. Next, consider $\Vec{d}$ along the $\hat{z}$-axis (middle column), which is always
described by case 2, and thus we see no splitting of the ABS for any angle $\phi_B$. Finally, we consider
a $\Vec{d}$ with an arbitrary orientation (right column), where the response is similar to the left column, with a key difference:
the maximum splitting is strictly less than $B$.
%{\bf Spatial structure of the order parameter:}
\section{Spatial structure of the order parameter}
Up to this point we have focused on elucidating the orbital structure
($\bm{k}$-dependence and nodes) and the spin structure ($\Vec{d}$-vector)
of the SC order parameter $\langle c_{\bm{k}\alpha}c_{-\bm{k}\beta} \rangle$.
While clearly uniform under the modified translations $\widetilde{T}(\bm{R})$, we ask whether it has interesting structure
under the physical translation ${T}(\bm{R})$ and on the graphene scale.

To address this question, we express $c_{\bm{k}\alpha}$ which annihilates an electronic excitation of the IKS metallic state,
in terms of the fermion operators $\psi_{\bm{k}\pm\bm{Q}/2,\tau,\alpha}$ where $\bm{Q}$ is the IKS wave-vector, the valley index
$\tau = K, K'$, and spin $\alpha = \uparrow,\downarrow$. (Following ref.~\cite{kwan2021}, we work
in the valley $\tau_z$ and ${\cal C} = \sigma_z\tau_z$ basis, with Chern number zero
bands that are equal superpositions of ${\cal C} = \pm 1$.)
The SC order parameter written in terms of the $\psi$'s has, amongst others, terms of the form
$\langle \psi_{\bm{k}+{\bm{Q}}/2,K,\alpha} \psi_{-\bm{k}+{\bm{Q}}/2,K,\beta} \rangle
+ \langle \psi_{\bm{k}-{\bm{Q}}/2,K',\alpha} \psi_{-\bm{k}-{\bm{Q}}/2,K',\beta} \rangle$;
see Supplementary Info.
These terms describe intra-valley pairing modulated at IKS wave-vector $\bm{Q}$.
Thus they break ${T}(\bm{R})$ while preserving $\widetilde{T}(\bm{R})$.

We must emphasize that this order parameter is {\it not} a pair density wave (PDW)~\cite{agterberg2020}.
In a PDW,  the pairing (off-diagonal) terms, with finite center-of-mass momentum, have a different $\bm{k}$-space structure relative to the kinetic energy (diagonal) terms.
In contrast, both the kinetic energy and pairing terms have the {\it same} momentum space structure in our problem.
%{\bf Experimental Implications:}
\section{Experimental Implications}
We conclude by summarizing our results focusing on experimental predictions.
Given a spin-rotation invariant IKS normal state, we can have either singlet or triplet pairing.
For triplet pairing, we further show that symmetry and nodal constraints imply that the direction of $\Vec{d}$ is $\bm{k}$-independent
for small hole doping away from $\nu\!=\!-2$.

A Zeeman field $\Vec{B}$ allows us to unambiguously discriminate between the allowed possibilities.
There are two general classes of such experiments for which we make predictions:
(a) those that probe low-energy Bogoliubov quasiparticle excitations in the bulk, and
(b) those that are sensitive to the phase of the order parameter and probe the low-energy Andreev bound states (ABS) near an interface.

Thermodynamic experiments that probe the near nodal excitations do not seem feasible, so we focus on tunneling spectroscopy.
For singlet pairing, the zero energy DOS $N(0)$ scales with $B$ independent of its in-plane direction.
When $\hat{d} = \hat{z}$, the low energy DOS $N(E)\sim |E|$, i.e., unaffected by $\Vec{B}$.
For an in-plane $\hat{d}$ the low energy DOS $N(E)$ shows a striking variation with the direction of $\Vec{B}$ relative to $\Vec{d}$ (see Fig.~1),
which is governed by how the Zeeman field impacts the nodal structure.

%Low-energy excitations near the nodes directly impact thermodynamic quantities, such as the
%low temperature specific heat (probably impossible to measure) and
%the spin susceptibility $\chi$, as described by eq.~\eqref{Eq:triplet-chi}.
%While a conventional NMR Knight shift measurement of $\chi$ may not be feasible in moire materials, perhaps it could be
%possible to measure $\chi$ using optical techniques.

Phase-sensitive measurements provide definitive proof of pairing symmetry, cf.~high $T_c$ cuprates. Our predictions for
ABS spectroscopy near an interface fall into this category. The interface between a insulator
and the superconductor can be gate-defined and does {\it not} have to be atomically sharp. The topologically
protected ABS are insensitive to the spatial variation of the potential normal to the interface (say, the $x$-direction).

While a smooth interface along $y$ underlies the topological protection of the ABS at $E=0$,
disorder will break $k_y$ conservation. However, it is important to note that
the ABS persist even in the presence of interfacial disorder~\cite{matsumoto1995,Fogrlstrom1997,Kalenkov2004},
except that they are shifted away from $E=0$ and may acquire a width.

The existence of ABS is robust for a sign-changing order parameter for almost all orientations of the
order parameter lobes with respect to the interface. Although there are very special orientations when ABS do not exist,
we do not expect this to be an issue, since experiments~\cite{Cao2021Nematicity,zhang2026} do not show any correlation between the nematicity
of the SC state and the underlying graphene lattice.

We note that ABS have been observed in tunneling spectroscopy in materials with unconventional pairing, such as the cuprates~\cite{Covington1996,Covington1997} and heavy fermions~\cite{Shrestha2021}. We have previously given compelling arguments~\cite{biswas2025} for STM tip-induced ABS as the explanation for
Andreev spectroscopy data in TBG~\cite{oh2021}. However, there are questions about the nature of the interface and its effect on the sample and its local density
when an STM tip is in contact with the sample. What we propose here are standard
STM experiments in the tunneling regime which probe the LDOS of the SC near an interface.
Since ABS exist all along the interface, a planar tunneling experiment that probes this region would also
have its low-energy spectrum dominated by the ABS.

The predictions for these measurements are shown in Fig.~3 for singlet pairing and in Fig.~4 for the triplet case.
For singlet pairing, we see that the $E\!=\!0$ spectral peak of the zero-field ABS is Zeeman split by $\pm B$,
independent of the in-plane orientation of $\Vec{B}$.
In the triplet case, when $\hat{d} = \hat{z}$, the $E\!=\!0$ is unaffected by a Zeeman field irrespective of its orientation.
While a SC with an in-plane $\Vec{d}$-vector gives rise to a characteristic variation in the ABS spectrum
as a function of the relative angle between the field and $\hat{d}$; see Fig.~3.
Taken together, these predictions would allow one to completely characterize the pairing symmetry in
moir\'e graphene SCs.

To conclude, we note that there are many directions in which the results presented above can be extended.
For instance, one can use our Green's function calculations to make predictions for
STM quasiparticle interference and for momentum-conserving tunneling in the quantum twist microscope~\cite{ilani2025},
which will probe the SC state in the near future.
Finally, the spatial structure of the order parameter on the graphene and moir\'e scales, briefly discussed in the preceding section,
leads only to intensity modulations of the spectral features described in our work, not to changes in the spectra.
It may be possible to probe the order parameter modulations using scanning Josephson spectroscopy~\cite{Hamidian2016},
an analysis of which is also left for the future.

\begin{acknowledgments}
We thank T. Hazra, E. Khalaf, H. Kim, S. Nadj-Perge, S. Parameswaran, T. Senthil, and J. Shan for discussions.
This work was supported in part by funds from The Ohio State University. RS acknowledges the support of the
Department of Atomic Energy, Government of India, under Project Identification No.~RTI 4012.
\end{acknowledgments}
\appendix
\section{Triplet $\Vec{d}$-vector}
\label{app:d_vector}
 We first derive the key result that the direction of the $\Vec{d}$-vector
is $\bm{k}$-independent. As noted in the text, fermion anticommutation and TR imply that $\Vec{d}(\bm{k})\!=\!-\Vec{d}(-\bm{k})\!=\!-\Vec{d}^*(-\bm{k})$.
Thus, each component $d_i(\bm{k})$ ($i=x,y,z$) is a real antisymmetric function of $\bm{k}$ which can be written as
$d_i(\bm{k}) = X_i k_x + Y_i k_y$  for small $k$. Writing $X_i = A_i\cos\beta_i$ and $Y_i = A_i\sin\beta_i$, we find
$d_i(\bm{k}) = A_i k\cos(\theta_{\bm{k}} - \beta_i)$.
Now we impose the constraint that the energy gap must vanish on the Fermi surface
along a nodal direction, which requires that all three components $d_i(\bm{k})$ vanish at the node.
This is possible if and only if $\beta_i = \beta$ is independent of $i$.
Thus
$\Vec{d}(\bm{k}) = k\cos(\theta_{\bm{k}} - \beta)\,(A_1,A_2,A_3)$, which is
of the form of eq.~\eqref{Eq:d-vector} with $\hat{d}$ independent of $\bm{k}$.

Next, we show there are only two allowed orientations for the $\Vec{d}$-vector -- in-plane or out-of-plane --
in the presence of a small SOC provided mirror $M_z$ is preserved.
Then $M_z$ transforms $d_{x,y}(\bm{k})\to -d_{x,y}(\bm{k})$  and $d_{z}(\bm{k})\to d_{z}(\bm{k})$.
Since symmetries are realized projectively~\cite{biswas2024} on the fermions, the order parameter, which is
a fermion bilinear, must transform under $M_z$ into itself up to a sign.
Thus $\hat{d}$-vector can only lie either in-plane or out-of-plane.
However, as emphasized in the main text, we will present results for arbitrary orientations of $\Vec{d}$.

\section{DOS in a Zeeman field}
\label{app:DOS}
We show here how $\Vec{B}$ impacts the near nodal dispersion in a triplet SC
and leads to the low-energy DOS shown in Fig.~1. We expand the spectrum $E_{\bm{k}}$ of eq.~\eqref{Eq:spectrum} in the vicinity of a
node using $\xi({\bm{k}}) \simeq v_F\delta k_\perp$ and $\Phi_1({\bm{k}})\simeq v_\Delta \delta k_\parallel$.
For simplicity, we have assumed here that the nodal direction of $\Phi_S(\bm{k})$ is perpendicular to the Fermi surface.
The more general case is discussed in the Supplementary Info, but the final results of interest are not impacted
in any essential way.

Consider the $\Vec{d}$-vector to lie in the plane and let $\Vec{B}$ be at an angle $\phi_B$ to $\hat{d}$, so that $B_\perp = B\sin\phi_B$.
To find the new nodes in the presence of a Zeeman field, we need only look at the negative sign expression in eq.~\eqref{Eq:spectrum}.
We see that $\phi_B = 0$ is special, and the original (zero-field) node is transformed into a small ellipse
$(v_F\delta k_\perp)^2 + (v_\Delta \delta k_\parallel)^2 = B^2$ of gapless excitations, a Bogoliubov Fermi surface [Fig.~1(b1)].
At any $\phi_B \neq 0$, there are only two (field-dependent) nodes [Fig.~1(b2,b3)]. The dispersion away from these new
nodes is gapped and one can analytically show that it has van Hove singularities (vHs) at $E = B\sin\phi_B$. These vHs lead to the low energy
``coherence peaks'' in the DOS, see Fig.~1.

We see that these coherence peaks are easily washed out by a small broadening $\Gamma$ introduced in
the $\delta$-functions of the low energy DOS $N(E) \sim \int d\delta k_\perp d\delta k_\parallel \delta(E - E_{\bm{k}})$.
Despite this, we see a characteristic change in the low energy DOS as a function of $\phi_B$ in Fig.~1.

%To analyze the dependence of the zero energy DOS $N(0)$ on the magnitude of the field, we can scale all electronic energies by $B$
%to obtain $N(0)\!\sim\!(B/v_F v_\Delta) F(\cos\phi_B, B/\Gamma)$, where the dimensionless
%$F = \int dx\, dy\, (\Gamma/B)/[x^2 +y^2 +1 - 2\sqrt{x^2\cos^2\phi_B +y^2} + (\Gamma/B)^2]$.
% We thus see that if we choose a $\Gamma$, when $B \ll \Gamma$, the effective broadening is large
% and $N(0)\!\sim\!(B/v_F v_\Delta) F(\cos\phi_B, 0)$.
% However, when $B \gg \Gamma$, the effective broadening is tiny, and the $\Gamma \to 0$ limit is
%singular, with $N(0)$ remaining finite for $\phi_B = 0$, but vanishing for any $\phi_B \neq 0$.
%\bigskip

%\noindent
\section{Green's functions and ABS}
\label{app:ABS}
We describe here how we compute the {\it half-space} Green's function (GF) $G(x,x^\prime, k_y; z)$ for $x\!>\!0, x^\prime\!>0$,
from whose spectral function we obtain the Andreev bound states (ABS) that exist near the boundary at $x=0$.
This underlies the methodology used in the calculating the results of Fig.~4.

As shown in the main text, the precise form of the boundary condition is not important since the
ABS are topologically protected, provided that the interface is smooth so that $k_y$ is conserved.
We will simply use a hard wall boundary at $x=0$, for which $G(x,x^\prime, k_y; z)$ is given by eq.~\eqref{Eq:half-space-GF}.

Our main task then is to compute the {\it translationally invariant, bulk} GF $G_0(x, k_y; z)$ in the presence of the Zeeman field, and use it
eq.~\eqref{Eq:half-space-GF} to find $G$. We focus on the $x$ or $k_x$ dependence of the retarded $G_0$ setting $z\!=E+i0^+$ and
suppress the $k_y$ and $E$ arguments, unless necessary.
The $\bm{k}$-space bulk $G_0$ was defined in the paragraph below eq.~\eqref{Eq:bdg1}.
We evaluate its Fourier transform $G_0(x)\!=\!\int_{-\infty}^{+\infty} (dk_x/2\pi)\,e^{ik_x x}\,{G}_0(k_x)$
using contour integration, and find
\begin{eqnarray}
G_0(x) &=& i\Theta(x) \sum_{k_0; {\rm Im}k_0 > 0}\,e^{ik_0 x}\,{\rm Res}\,{G}_0(k_0)
\label{Eq:r-space-GF} \\
&-& i\Theta(-x) \sum_{k_0; {\rm Im}k_0 < 0}\,e^{-ik_0 x}\,{\rm Res}\,{G}_0(k_0), \nonumber
\end{eqnarray}
where the sums are over the poles of ${G}_0(k_x)$ at $k_x = k_0$ and ``Res'' denotes the corresponding residues.

Since we have established the topological protection of the ABS for smooth interfaces (where $k_y$ is conserved),
the results for the ABS are valid over the entire BCS-side of the crossover. This allows us to do the calculation in the weak coupling BCS limit where
the analysis is the simplest. It turns out that for fixed $k_y$ and $E$, the bulk GF ${G}_0(k_x)$ has eight poles in the complex $k_x$-plane in the presence of a Zeeman field.
Four of the poles are in the upper-half plane (${\rm Im}\, k_0 > 0$) and four in the lower-half (${\rm Im}\,k_0 < 0$).
We find explicit algebraic expressions for the location of the 8 poles and the corresponding residues.
The (tedious) details are relegated to the Supplementary Info. Using these results in eq.~\eqref{Eq:r-space-GF}
we determine $G_0(x)$, which in turn leads to $G(x,x^\prime)$ via eq.~\eqref{Eq:half-space-GF}.
\bigskip
\bibliographystyle{apsrev4-2}
\bibliography{pairingsymmetry.bib}

\clearpage
\onecolumngrid  % switch the remainder of the document (the SM) to single-column

\beginsupplement

\begin{center}
\textbf{\large Supplemental Material}
\end{center}

\vspace{1cm}

\section*{I. Review of the symmetries of the IKS normal state}
\label{sec:Normal_state}
For completeness, in this section we review the incommensurate Kekul\'e spiral (IKS) normal state at fillings $-3<\nu_c<\nu<-2$ and its symmetries, following the pioneering works of Refs.~\cite{kwan2021,wagner2022,wang2025}. We discuss: (1) the effective moir\'e translation $\Tilde{T}(\bm{R})$, which combines an ordinary moir\'e translation with a valley-dependent gauge transformation; (2) the twofold rotation about the $z$ axis, $\mathcal{C}_{2z}$; (3) time-reversal symmetry, $\mathcal{T}$; and (4) global $SU(2)$ spin rotations. In particular, we establish the symmetry representations of $\mathcal{C}_{2z}$ and $\mathcal{T}$ in the spin-degenerate IKS degrees of freedom from which superconductivity arises, as defined immediately above Eq.~[1] in the main text.

In terms of $\psi^\dagger_{\bm{k}a}$ $\bigl(\psi_{\bm{k}a}\bigr)$, the creation (annihilation) operators for the eight moir\'e flat-band states labeled by $a=(\tau,\sigma,\alpha)$, where $\tau= K,K^\prime$ denotes the valley index, $\sigma=A,B$ the sublattice (band) index, and $\alpha= \uparrow,\downarrow$ the spin index, the IKS order is characterized by
\begin{equation}
    \bigl\langle
    \psi^{\dagger}_{\bm{k}_1,K,\sigma_1,\alpha}
    \psi_{\bm{k}_2,K^\prime,\sigma_2,\beta}
    \bigr\rangle
    \propto
    \delta_{\bm{k}_1-\bm{k}_2,\bm{Q}},
    \label{eq:supp:IKSorder}
\end{equation}
where we elaborate the sublattice and spin structure below. This state exhibits intervalley coherence (IVC), corresponding to spontaneous breaking of the valley $U(1)$ symmetry, which manifests as a Kekul\'e pattern on the graphene scale. In contrast to the homogeneous IVC order~\cite{bultinck2020} stabilized in the absence of strain, the strain stabilized IKS state exhibits an IVC order that \textit{spirals} with an \textit{incommensurate} wave vector $\bm{Q}$ on the moir\'e scale, thereby breaking the ordinary translation symmetry by moir\'e lattice vector $\bm{R}$, $T(\bm{R})$, under which
\begin{equation}
    T(\bm{R}):\quad
    \psi_{\bm{k}}
    \rightarrow
    e^{-i\bm{k}\cdot\bm{R}}
    \psi_{\bm{k}}.
\end{equation}
It follows immediately from Eq.~\eqref{eq:supp:IKSorder} that the IKS state is instead invariant under the modified moir\'e translation
\begin{equation}
    \Tilde{T}(\bm{R}):\quad
    \psi_{\bm{k}}
    \rightarrow
    e^{-i\bm{k}\cdot\bm{R}}
    e^{i\bm{Q}\cdot\bm{R}\tau_z/2}
    \psi_{\bm{k}}.
    \label{eq:supp:modified_translation}
\end{equation}

Moreover, within Hartree--Fock (HF) theory at the relevant fillings, the inclusion of intervalley scattering stabilizes an IKS state with \textit{global} $SU(2)$ spin-rotation symmetry out of a manifold of degenerate states related by $SU(2)_K \times SU(2)_{K^\prime}$. This state is predicted to exhibit only a charge Kekul\'e pattern, with no spin Kekul\'e pattern, in agreement with experiment. Furthermore, the IKS state preserves both the twofold rotation symmetry $\mathcal{C}_{2z}$ and time-reversal symmetry $\mathcal{T}$:
\begin{equation}
\begin{aligned}
\mathcal{C}_{2z}: \quad &
\psi_{\bm{k}}
\rightarrow
\sigma_x\tau_x\psi_{-\bm{k}},\\
\mathcal{T}: \quad &
\psi_{\bm{k}}
\rightarrow
i\tau_x s_y\psi_{-\bm{k}}.
\end{aligned}
\label{eq:supp:C2_and_T}
\end{equation}
At this stage, it is convenient to define
\begin{equation}
    \Tilde{\psi}_{\bm{k}}
    =
    \psi_{\bm{k}+\tau_z\bm{Q}/2},
    \label{eq:supp:psi_tilde}
\end{equation}
which transforms as
\begin{equation}
    \Tilde{T}(\bm{R}):~\Tilde{\psi}_{\bm{k}}
    \rightarrow
    e^{-i\bm{k}\cdot\bm{R}}
    \Tilde{\psi}_{\bm{k}},
\end{equation}
and is therefore labeled by the eigenvalue of $\Tilde{T}(\bm{R})$, which is a good quantum number. Combining Eqs.~\eqref{eq:supp:C2_and_T} and \eqref{eq:supp:psi_tilde}, we find that the representations of $\mathcal{C}_{2z}$ and $\mathcal{T}$ acting on $\Tilde{\psi}_{\bm{k}}$ are unchanged, with $\mathcal{C}_{2z}: \quad
\Tilde{\psi}_{\bm{k}}
\rightarrow
\sigma_x\tau_x\Tilde{\psi}_{-\bm{k}}$ and $\mathcal{T}: \quad
\Tilde{\psi}_{\bm{k}}
\rightarrow
i\tau_x s_y\Tilde{\psi}_{-\bm{k}}$.

In terms of $\Tilde{\psi}_{\bm{k}}$, the single-particle density matrix,
\begin{equation}
    \bigl\langle
    \Tilde{\psi}_{\bm{k}\tau\sigma\alpha}^{\dagger}
    \Tilde{\psi}_{\bm{k}'\tau'\sigma'\beta}
    \bigr\rangle
    =
    \bigl[P(\bm{k})\bigr]_{\tau\sigma,\tau'\sigma'}
    \delta_{\alpha\beta}
    \delta_{\bm{k}\bm{k}'},
\end{equation}
is diagonal in both momentum and spin. The $4\times4$ matrix in valley--sublattice space, $P(\bm{k})$, satisfies the filling constraint $2A_{\mathrm{M}}\int_{\bm{k}\in\mathrm{mBZ}}\operatorname{Tr}\!\left[P(\bm{k})\right]=4+\nu$, where $A_{\mathrm{M}}$ is the moir\'e unit-cell area. Within HF theory, the order parameter $P(\bm{k})$ is a projection operator, i.e., $P^2(\bm{k})=P(\bm{k})$. As shown in \cite{wang2025}, the hole-doped state at $-3<\nu_c<\nu< -2$ inherits the symmetries of the $\nu=-2$ insulator, it suffices, for the purpose of symmetry analysis, to consider the $\nu=-2$ state. In this case, $P(\bm{k})$ is a rank-one projector, i.e., $\operatorname{Tr}P(\bm{k})=1$ at every $\bm{k}$ in the moir\'e Brillouin zone.

The $4\times4$ $\tau$--$\sigma$ structure of $P(\bm{k})$ is most transparently encoded by introducing two mutually commuting $SU(2)$ Lie algebras, $\vec{\gamma}
    =
    \bigl(\sigma_x,\tau_z\sigma_y,\tau_z\sigma_z\bigr),~
    \vec{\eta}
    =
    \bigl(\tau_x\sigma_x,\tau_y\sigma_x,\tau_z\bigr)$
where $\gamma_z=\tau_z\sigma_z$ is the Chern number operator. The HF projector then takes the factorized form (see Eq.~(6) of Ref.~\cite{kwan2021})
\begin{equation}
    P(\bm{k})
    =
    \frac{1}{4}
    \bigl(\mathbb{I}+\hat{n}_{\bm{k}}\cdot\vec{\gamma}\bigr)
    \bigl(\mathbb{I}+\hat{m}_{\bm{k}}\cdot\vec{\eta}\bigr),
\end{equation}
where
$\hat{n}_{\bm{k}}
=
(\cos\phi_{\bm{k}},\sin\phi_{\bm{k}},0)$
and
$\hat{m}_{\bm{k}}
=
(\sin\theta_{\bm{k}},0,\cos\theta_{\bm{k}})$.
Invariance under $\mathcal{T}$ and $\mathcal{C}_{2z}$ requires
$\phi_{-\bm{k}}=\phi_{\bm{k}}$
and
$\theta_{-\bm{k}}=\pi-\theta_{\bm{k}}$.
The confinement of $\hat{n}_{\bm{k}}$ to the $xy$ plane enforces equal-weight mixing of the two opposite Chern sectors ($\gamma_z=\pm1$), yielding a Chern-zero state.
% The residual gauge freedom $e^{i\Phi_{\bm{k}}\gamma_z}$ may then be exploited to set $\phi_{\bm{k}}=0$.

At $\nu=-2$, the lowest (valence) band, which is fully occupied and energetically separated from the conduction band(s), corresponds to the $+1$ eigenstate of $P(\bm{k})$. In terms of $\Tilde{\psi}_{\bm{k}}$, its annihilation operator is
\begin{equation}
    c_{\bm{k}\alpha}
    =
    \cos\!\Bigl(\frac{\theta_{\bm{k}}}{2}\Bigr)
    \displaystyle\psi_{\bm{k}+\bm{Q}/2,K,\alpha}
    +
    \sin\!\Bigl(\frac{\theta_{\bm{k}}}{2}\Bigr)
    \psi_{\bm{k}-\bm{Q}/2,K^\prime,\alpha},
    \label{eq:supp:IKS_band}
\end{equation}
where $\psi_{\bm{k}+\bm{Q}/2,K,\alpha} = \Bigl(
      e^{i\frac{\phi_k}{2}}\,\Tilde{\psi}_{\bm{k}K,A,\alpha}
      +
      e^{-i\frac{\phi_k}{2}}\,\Tilde{\psi}_{\bm{k}K,B,\alpha}\Bigr)/
      {\sqrt{2}}
      $ and $\psi_{\bm{k}-\bm{Q}/2,K^\prime,\alpha}=\Bigl(
      e^{-i\frac{\phi_k}{2}}\,\Tilde{\psi}_{\bm{k}K^\prime,A,\alpha}
      +
      e^{i\frac{\phi_k}{2}}\,\Tilde{\psi}_{\bm{k}K^\prime,B,\alpha}\Bigr)/
      {\sqrt{2}}
      $.
Using these results together with
$\sin\!\bigl(\theta_{\bm{k}}/2\bigr)
=
\cos\!\bigl(\theta_{-\bm{k}}/2\bigr)$
and
$\phi_{\bm{k}}
=
\phi_{-\bm{k}}$,
we obtain the representations of $\mathcal{C}_{2z}$ and $\mathcal{T}$ on the spin-degenerate IKS band,
\begin{equation}
    \mathcal{C}_{2z}:\;
    c_{\bm{k}\alpha}
    \rightarrow
    c_{-\bm{k}\alpha},
    \qquad
    \mathcal{T}:\;
    c_{\bm{k}\alpha}
    \rightarrow
    (is_y)_{\alpha\beta}
    c_{-\bm{k}\beta}.
\end{equation}

\section*{II. Gap nodes and density of states}

In the main text, we showed that a small Fermi pocket around the $\Gamma$ point ($\bm{k}=0$), together with time-reversal symmetry and a nodal order parameter, implies that, to leading order, $\Vec{d}(\bm{k})\approx\Phi_1(\bm{k})\,\hat{d}$, where $\Phi_1(\bm{k})$ is a $p$-wave form factor and $\hat{d}$ is independent of $\bm{k}$. In our analysis of the low-energy density of states (DOS), we made the simplifying assumption that, at a gap node $\bm{k}_0$, the direction along which $\Vec{d}(\bm{k})$ vanishes (the nodal direction) coincides with the normal to the Fermi surface. This is the case when the node lies at a high-symmetry point on the Fermi surface. Expanding about $\bm{k}_0$ in terms of momenta normal ($\delta k_\perp$) and tangential ($\delta k_\parallel$) to the Fermi surface, we obtained $\Phi_1(\bm{k}_0+\delta\bm{k})=v_\Delta\,\delta k_\parallel$, together with $\xi(\bm{k}_0+\delta\bm{k})\approx v_F\,\delta k_\perp$.

In what follows, we relax these assumptions. We consider an expansion about a generic nodal point $\bm{k}_0$, which need not lie near the $\Gamma$ point, and allow the nodal direction to be oriented arbitrarily in the $(\delta k_\parallel,\delta k_\perp)$ plane. Specifically, we assume that, to linear order in $\delta\bm{k}$, $\Vec{d}(\bm{k}_0+\delta\bm{k})=0$ when $\delta \bm{k}$ lies along the nodal line which we choose to be at an angle $-\,\gamma$ with respect to the Fermi surface normal (the negative sign is for later algebraic convenience). This implies that
\begin{eqnarray}
    \Vec{d}(\bm{k})
    \approx \Phi_1(\bm{k})\,\hat{d}
    = v_\Delta\bigl(\delta k_\parallel\cos\gamma
    + \delta k_\perp\sin\gamma\bigr)\hat{d}.
\end{eqnarray}
For $\gamma=0$, we recover the results discussed in the main text. We show below that our conclusions regarding the low-energy DOS remain unchanged in this more general case.

In the absence of a Zeeman field, the quasiparticle dispersion in the vicinity of $\bm{k}_0$, given by $E_0^2(\bm{k})=\xi^2(\bm{k})+\Phi_1^2(\bm{k})$, has a Dirac node at $\delta k_\parallel=\delta k_\perp=0$. A Zeeman field $\Vec{B}$, modifies the dispersion to
\begin{subequations}
    \begin{eqnarray}
        E_{\pm}^2\left(\bm{k}\right) &=& \xi^2\left(\bm{k}\right)+ \Phi^2_1\left(\bm{k}\right) + B^2 \pm 2\sqrt{B^2\xi^2\left(\bm{k}\right)+ B^2_{\parallel}\Phi^2_1\left(\bm{k}\right)}\\
        &=& E^2_0\left(\bm{k}\right) + B^2 \pm 2\sqrt{B^2E_0^2\left(\bm{k}\right)- B^2_{\perp}\Phi^2_1\left(\bm{k}\right)}
    \end{eqnarray}
    \label{eq:supp_spectrum}
\end{subequations}
where $\Vec{B}_{\parallel} = \left(\Vec{B}\cdot\hat{d}\right)\hat{d}$ and $\Vec{B}_{\perp} = \Vec{B} - \left(\Vec{B}\cdot\hat{d}\right)\hat{d}$. As we discuss below, the locus of zero-energy excitations in $\bm{k}$ space, determined by
\begin{equation}
    E^2_0\left(\bm{k}\right) + B^2 - 2\sqrt{B^2E_0^2\left(\bm{k}\right)- B^2_{\perp}\Phi^2_1\left(\bm{k}\right)} = 0,
    \label{eq:supp:Ek=0}
\end{equation}
is no longer a single node at $\delta\bm{k}=0$. We study its evolution as a function of $\phi_B$, the angle between $\Vec{B}$ and $\hat{d}$.

\begin{figure}[t]
    \centering
    \includegraphics[width=\textwidth]{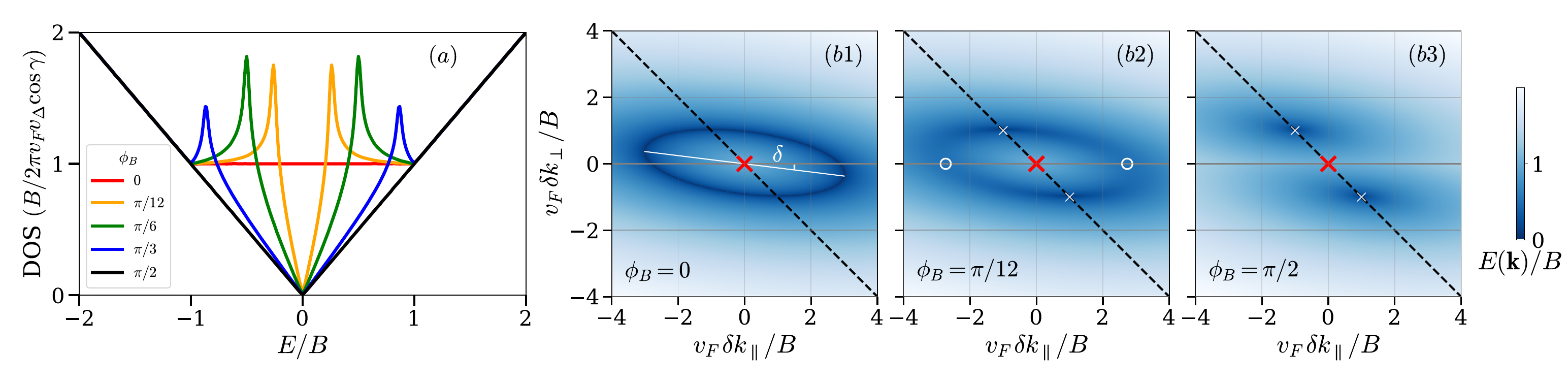}
    \caption{(a) Density of states for several values of $\phi_B$, with $v_\Delta/v_F =0.5$ and $\gamma=\pi/4$. (b1)--(b3) False-color plots of $E_-(\mathbf{k})$ in the $\left( \delta k_\parallel,\, \delta k_\perp\right)$ plane for $\phi_B=0,\,\pi/12,\,\pi/2$, respectively. The red cross marks the location of the node at $B=0$, the dashed black line is the nodal direction, and in (b1) the angle $\delta$ between the $k_\parallel$ axis and the principal axis of the elliptical contour is indicated. Open circles in (b2) mark the saddle points giving rise to Van-Hove sigularities in the DOS.}
    \label{fig:S1}
\end{figure}

A more physically relevant quantity is the low-energy density of states (DOS),
\begin{equation}
    N\left(E\right) = \int_{\bm{k}}\,\Bigl( \delta\left(\left|E\right|- E_+\left(\bm{k}\right)\right) + \delta\left(\left|E\right|- E_-\left(\bm{k}\right)\right)\Bigr),
    \label{eq:supp:DOS_def}
\end{equation}
where $\int_{\bm{k}} := \int d\delta k_\parallel d \delta k_\perp/\left(2\pi\right)^2$. At $B=0$, $E_\pm(\bm{k})=E_0(\bm{k})$, and the Dirac spectrum yields a DOS that is linear in energy,
\begin{equation}
    N_{0}\left(E\right) = 2\int_{\bm{k}} \delta\left(\left|E\right|- E_0\left(\bm{k}\right)\right) = \frac{\left|E\right|}{\pi v_Fv_\Delta \cos\gamma}.
    \label{eq:supp:DOS_B=0}
\end{equation}
We now examine how Eq.~\eqref{eq:supp:DOS_B=0} is modified for $B\neq0$ and how it evolves as a function of $\phi_B$. The results are summarized in Fig.~\ref{fig:S1}(a).

\bigskip
$\Vec{\mathbf{B}}\parallel\hat{\mathbf{d}}:\ $In this case $B_\perp=B\sin\phi_B=0$. Equation~\eqref{eq:supp:Ek=0} is then satisfied for $E_0^2(\bm{k})=B^2$. The corresponding locus,
\begin{equation}
    \bigl(v_F^2+v_\Delta^2\sin^2\gamma\bigr)\delta k_\perp^2
    +v_\Delta^2\sin2\gamma\,\delta k_\parallel\delta k_\perp
    +v_\Delta^2\cos^2\gamma\,\delta k_\parallel^2
    =B^2,
    \label{eq:supp:ellipse}
\end{equation}
describes an ellipse with major and minor semiaxes $A_+$ and $A_-$, respectively, given by
\begin{equation}
    \frac{1}{A_\pm^2}
    =
    \frac{v_F^2}{2B^2}
    \Biggl[
    \biggl(1+\frac{v_\Delta^2}{v_F^2}\biggr)
    \mp
    \sqrt{
    \biggl(1-\frac{v_\Delta^2}{v_F^2}\cos2\gamma\biggr)^2
    +
    \biggl(\frac{v_\Delta^2}{v_F^2}\sin2\gamma\biggr)^2
    }
    \Biggr].
\end{equation}
When the nodal direction is normal to the Fermi surface, i.e., $\gamma=0$, and $v_\Delta<v_F$, the major (minor) axis lies along the $\delta k_\parallel$ ($\delta k_\perp$) direction and has length $B/v_\Delta$ ($B/v_F$). For $\gamma\neq0$, the major axis is rotated by an angle
\begin{equation}
    \delta
    =
    \frac{1}{2}
    \tan^{-1}
    \Bigl(
    \frac{\sin2\gamma}
    {v_F^2/v_\Delta^2-\cos2\gamma}
    \Bigr)
\end{equation}
with respect to the $\delta k_\parallel$ axis. For $v_\Delta<v_F$, $\delta$ attains its maximum value,
$\frac{1}{2}\tan^{-1}(v_\Delta^2/v_F^2)$, at $\gamma=\pi/4$. The low-energy excitation spectrum for $\Vec{B}\parallel\hat{d}$ is shown in Fig.~\ref{fig:S1}(b1) for $\gamma=\pi/4$, clearly exhibiting the ellipse of zero-energy excitations. The nodal direction is indicated by the black dashed line, while the major axis of the ellipse, tilted by an angle $\delta$ with respect to the horizontal axis, is shown in white.
\bigskip

\textbf{DOS for} $\Vec{\mathbf{B}}\parallel\hat{\mathbf{d}}.$ Setting $B_\perp=0$ in Eq.~\eqref{eq:supp_spectrum}, we obtain $E_\pm(\bm{k})=\left|E_0(\bm{k})\pm B\right|$, and
\begin{subequations}
\begin{align}
N(E)
&=
\int_{\bm{k}}
\Bigl(\delta\bigl(|E|-|E_0(\bm{k})+B|\bigr)
+
\delta\bigl(|E|-|E_0(\bm{k})-B|\bigr)\Bigr)
\\
&=
\int_{\bm{k}}
\delta\bigl(|E|-E_0(\bm{k})-B\bigr)
+
\int_{\bm{k};\,E_0(\bm{k})<B}
\delta\bigl(|E|+E_0(\bm{k})-B\bigr)
\nonumber\\
&\quad\qquad\qquad\qquad\qquad\qquad
+
\int_{\bm{k};\,E_0(\bm{k})>B}
\delta\bigl(|E|-E_0(\bm{k})+B\bigr).
\label{eq:supp_DOS_phi=0_1}
\end{align}
\end{subequations}
For $\left|E\right|\geq B$, the first and third terms in Eq.~\eqref{eq:supp_DOS_phi=0_1} contribute, whereas for $\left|E\right|<B$, the second and third terms contribute, yielding
\begin{equation}
    N(E)=
    \begin{cases}
        \dfrac{B}{\pi v_F v_\Delta \cos\gamma }  & \text{if}\ \left|E\right|<B,\\[6pt]
        \dfrac{\left|E\right|}{\pi v_F v_\Delta \cos\gamma }  & \text{if}\ \left|E\right|\geq B,
    \end{cases}
    \qquad
    = \frac{1}{2}\bigl(N_0(E+B)+N_0(E-B)\bigr).
\end{equation}
The DOS is plotted in red in Fig.~\ref{fig:S1}(a). The finite zero-energy DOS, $N(0)\propto B$, is consistent with the presence of a Bogoliubov ``Fermi surface'' whose characteristic size is proportional to $B$.

\bigskip
$\Vec{\mathbf{B}}\nparallel \hat{\mathbf{d}}.$ For $\phi_B\neq0$ ($B_\perp\neq0$), Eq.~\eqref{eq:supp:Ek=0} is satisfied only when $\Phi_1(\bm{k})=0$ and $\xi^2(\bm{k})=B^2$. In other words, all points on the ellipse are gapped except at $\pm\bm{k}_*$, where the nodal line intersects the ellipse. The corresponding momenta are
\begin{equation}
    \bm{k}_*=\left(-\tan\gamma\,\frac{B}{v_F},\,\frac{B}{v_F}\right).
\end{equation}
In particular, for $\phi_B=\pi/2$, setting $B_\perp=B$ in Eq.~\eqref{eq:supp_spectrum} gives
\[
E_\pm(\bm{k})=\sqrt{\bigl(\xi(\bm{k})\pm B\bigr)^2+\Phi_1^2(\bm{k})}
            =E_0(\bm{k}\pm\bm{k}_0)
\]
for $\xi(\bm{k})>0$ ($\delta k_\perp>0$), and
\[
E_\pm(\bm{k})=\sqrt{\bigl(\xi(\bm{k})\mp B\bigr)^2+\Phi_1^2(\bm{k})}
            =E_0(\bm{k}\mp\bm{k}_0)
\]
for $\xi(\bm{k})<0$ ($\delta k_\perp<0$). Thus, the spectrum can be viewed as two copies of the $B=0$ Dirac cone centered at $\pm\bm{k}_0$. Consequently,
\begin{subequations}
    \begin{eqnarray}
       N\left(E\right) &=& \int_{\bm{k}} \delta\bigl(\left|E\right|-E_0\left(\bm{k}+\bm{k}_0\right)\bigr) + \int_{\bm{k}} \delta\bigl(\left|E\right|-E_0\left(\bm{k}-\bm{k}_0\right)\bigr) \\
        &=& 2\int_{\bm{k}} \delta\bigl(\left|E\right|-E_0\left(\bm{k}\right)\bigr) = N_0\left(E\right),
        \label{eq:supp_DOS_phi=pi/2}
    \end{eqnarray}
\end{subequations}
which is identical to the DOS at $B=0$, as shown by the black curve in Fig.~\ref{fig:S1}(a). For $0<\phi_B<\pi/2$, $N(0)=0$, consistent with the presence of two Dirac nodes at $\pm\bm{k}_*$. The van Hove singularities at $|E|=B\sin\phi_B$ originate from the saddle points $\pm\bar{\bm{k}}=\pm\left(B\cos\phi_B/(v_\Delta\cos\gamma),\,0\right)$ of $E_-(\bm{k})$ (shown by the open white circles in Fig.~\ref{fig:S1}(b2)), where $E_-(\bar{\bm{k}})=B\sin\phi_B$.

\section*{III. Andreev bound state spectroscopy}
\label{sec:ABS_spectroscopy}
In this section, we outline the algebraic steps involved in constructing the half-space Green's function (GF) $G(x,x^\prime)$ from the bulk  GF $G_0(x)$. The relationship between the two for a hard wall boundary, where $G(0,x^\prime)=G(x,0)=0$, is given by
\begin{equation}
    G\left(x,x^\prime\right)
    =
    G_0\left(x-x^\prime\right)
    -
    G_0(x)
    G_0(0)^{-1}
    G_0(-x^\prime).
    \label{eq:supp:half_space_1}
\end{equation}
as shown in the main text. We will compute $G(x,x^\prime)$ and
explicitly demonstrate the existence of zero-energy Andreev bound states (ABS)
by analyzing the density of states in the vicinity of the boundary.
Here we focus on the $B=0$ case, and look at the effects of a Zeeman field in the next Section.

The ABS are topologically protected at $E=0$ for a $k_y$ conserving interface, as discussed in detail in the main text.
This makes the answer insensitive to the choice of boundary condition and the hard wall boundary condition
that we use is only for theoretical convenience. Topological protection further guaranties $E=0$ ABS for
the entire weak pairing (BCS like) regime. This allows us to calculate in the $k_F\xi_0 \gg 1$ limit for theoretical ease.

As indicated in Appendix III, our approach is to first compute the bulk GF in real space
\begin{equation}
    G_0\left(x,k_y,z\right)
    =
    \int \frac{dk_x}{2\pi}
    e^{ik_xx}
    G_0\left(\bm{k},z\right).
\end{equation}
Unless necessary, we suppress the dependence on $k_y$ and complex energy $z$ and focus on the variables $x$ and $k_x$. The above integral is evaluated by contour integration, closing the contour in the upper (lower) half of the complex $k_x$ plane for $x>0$ ($x<0$). This yields
\begin{eqnarray}
G_0(x)
&=&
i\Theta(x)
\sum_{k_0;\,{\rm Im}\,k_0>0}
e^{ik_0x}\,
{\rm Res}\,{G}_0(k_0)
\nonumber\\
&&
-i\Theta(-x)
\sum_{k_0;\,{\rm Im}\,k_0<0}
e^{-ik_0x}\,
{\rm Res}\,{G}_0(k_0),
\label{eq:supp:r-space-GF}
\end{eqnarray}
where $k_0$ are the poles of $G_0(k_x)$.

In the absence of a Zeeman field the Hamiltonian for both the singlet $\left(S=0\right)$ and triplet $\left(S=1\right)$ cases can be written as
\begin{equation}
\mathcal{H}\left(\bm{k}\right) = \bigl(\xi\left(\bm{k}\right)\tau_z + \Phi_1\left(\bm{k}\right)\tau_x\bigr)\otimes\sigma_0 := \bm{u}(\bm{k})\cdot\bm{\tau}\otimes \sigma_0\label{eq:supp:ham_B=0}
\end{equation}
in an appropriately chosen basis; see main text.
The bulk GF is then given by
\begin{eqnarray}
    G_0\left(\bm{k},z\right) &=& \bigl(z\mathbb{I}-\mathcal{H}\left(\bm{k}\right)\bigr)^{-1}\nonumber\\
        &=& \frac{z\tau_0+\xi\left(\bm{k}\right)\tau_z + \Phi_1\left(\bm{k}\right)\tau_x}{z^2-\xi^2\left(\bm{k}\right)-\Phi^2_S\left(\bm{k}\right)}\otimes\sigma_0:=g_0\left(\bm{k},z\right)\otimes \sigma_0\label{eq:supp:GF_bulk}
\end{eqnarray}
At fixed $k_y$ and $z=E+i0^+$ we evaluate $G_0(x)$ using eq.~\eqref{eq:supp:r-space-GF}. To this end we solve for the poles of $G_0\left(k_x\right)$ by setting the denominator of \eqref{eq:supp:GF_bulk} to zero, which gives
\begin{equation}
    \xi^2\left(k_x\right) = \left(E+i0^+\right)^2 - \Phi^2_S\left(k_x\right).
\label{eq:supp:pole_condition}
\end{equation}
Suppose $\xi\left(k_x\right)$, the effective 1D dispersion at given $k_y$, has Fermi points $\pm k_{Fx}$ where $\xi(\pm k_{Fx})=0$ and $\partial_{k_x}\xi(\pm k_{Fx})=\pm v_{Fx}$ and suppose $\Phi_S(\pm k_{Fx}) = \Delta_\pm$. In the extreme weak coupling limit, $\left|\Delta_\pm\right|/v_{Fx}k_{Fx}\ll 1$, solving eq.~\eqref{eq:supp:pole_condition} yields two poles near $+k_{Fx}$ and two near $-k_{Fx}$ which we obtain by approximating
\begin{subequations}
    \begin{eqnarray}
        \xi\left(\pm k_{Fx} + \delta k_x\right) &\approx& \pm v_{Fx}\delta k_x\\
        \Phi_S\left(\pm k_{Fx} + \delta k_x\right) &\approx& \Delta_\pm
    \end{eqnarray}
    \label{eq:supp:linear_dispersion}
\end{subequations}
in eq.~\eqref{eq:supp:pole_condition}. We emphasize that the Fermi momentum $k_{Fx}$, Fermi velocity $v_{Fx}$ and $\Delta_\pm$ of the effective 1D problem are implicitly $k_y$ dependent. Furthermore, for simplicity we work in the situation where Fermi points and velocities are the same for left and right movers. Generalization to the case where left and right movers have different velocities is straightforward and does not lead to any qualitatively different result. However, in general $\Delta_+ \neq \Delta_-$ is essential to capture ABS. The four poles and the corresponding residues given by
\begin{equation}
    i~\text{Res}\,G_0\left(k_0\right) = -i~\frac{\left(E+i0^+\right)\tau_0+\xi\left(k_0\right)\tau_z + \Phi_1\left(k_0\right)\tau_x}{2\xi\left(k_0\right)\left(\partial \xi/\partial k_x\right)_{k_0}}\otimes\sigma_0
\end{equation}
are listed in table~\ref{tab:residues1}. The symbols used are defined in the caption.
\begin{table}[t]
\centering
\renewcommand{\arraystretch}{1.8}
\begin{tabular}{c@{\hspace{2cm}}c}
\toprule
$k_0$ & $iv_{Fx}\,\mathrm{Res}\,G_0(k_0)$ \\
\midrule

\multicolumn{2}{c}{$\mathrm{Im}(k_0)>0$} \\
\midrule

$\displaystyle k_{1+} = +k_{Fx}+\frac{\zeta_{+}}{v_{Fx}}$
&
$\displaystyle
\mathcal{P}_{+}
=
-i\frac{\mathcal{M}\!\left(\zeta_{+},\Delta_+\right)}
{2\zeta_{+}}
$
\\[2ex]

$\displaystyle k_{1-} =-k_{Fx}+\frac{\zeta_{-}}{v_{Fx}}$
&
$\displaystyle
\mathcal{P}_{-}
=
-i\frac{\mathcal{M}\!\left(-\zeta_{-},\Delta_-\right)}
{2\zeta_{-}}
$
\\[2ex]

\midrule

\multicolumn{2}{c}{$\mathrm{Im}(k_0)<0$} \\
\midrule

$\displaystyle k_{2+} = +k_{Fx}-\frac{\zeta_{+}}{v_{Fx}}$
&
$\displaystyle
-\mathcal{Q}_{+}
=
i\frac{\mathcal{M}\!\left(-\zeta_{+},\Delta_+\right)}
{2\zeta_{+}}
$
\\[2ex]

$\displaystyle k_{2-} = -k_{Fx}-\frac{\zeta_{-}}{v_{Fx}}$
&
$\displaystyle
-\mathcal{Q}_{-}
=
i\frac{\mathcal{M}\!\left(\zeta_{-},\Delta_-\right)}
{2\zeta_{-}}
$
\\

\bottomrule
\end{tabular}
\caption{Poles of the Green's function and the corresponding residues. For compactness we have defined $\zeta_{\pm} = \text{sgn}\left(E\right)\sqrt{\left(E+i0^+\right)^2-\Delta^2_\pm}$ such that $\text{Im}\left(\zeta_{\pm}\right)>0$ and $\mathcal{M}\left(\zeta,\Delta\right) = \bigl(\left(E+i0^+\right)\tau_0 + \zeta\tau_z + \Delta\tau_x\bigr)\otimes\sigma_0$.}
\label{tab:residues1}
\end{table}
Thus,
\begin{equation}
    v_{Fx}G_0\left(x\right) = \Theta\left(x\right)\biggl(e^{ik_{1+}x} \mathcal{P}_{+} + e^{ik_{1-}x} \mathcal{P}_{-}\biggr) + \Theta\left(-x\right)\biggl(e^{ik_{2+}x} \mathcal{Q}_{+} + e^{ik_{2-}x} \mathcal{Q}_{-}\biggr)
    \equiv v_{Fx}g_0(x)\otimes \sigma_0.
    \label{eq:supp:G0(x)}
\end{equation}
From Eq.~\eqref{eq:supp:G0(x)} and Table~\ref{tab:residues1}, it follows that
$v_{Fx}G_0(x\to0^+)=\mathcal{P}_+ + \mathcal{P}_-$ and
$v_{Fx}G_0(x\to0^-)=\mathcal{Q}_+ + \mathcal{Q}_-$ are identical, yielding
\begin{equation}
    v_{Fx}G_0(0) = -\frac{i}{2}\Biggl((E+i0^+)\bigl(\zeta^{-1}_+ + \zeta^{-1}_-\bigr)\tau_0 + \bigl(\Delta_+\zeta^{-1}_+ + \Delta_-\zeta^{-1}_-\bigr)\tau_x\Biggr)\otimes\sigma_0.
    \label{eq:supp:G0(0)}
\end{equation}
Using Eqs.~\eqref{eq:supp:half_space_1}, \eqref{eq:supp:G0(x)}, and \eqref{eq:supp:G0(0)}, we obtain the half-space GF
$G(x,x^\prime)\equiv g(x,x^\prime)\otimes\sigma_0$, where
\begin{equation}
g(x,x^\prime)
=
g_0(x-x^\prime)
-
g_0(x)g_0(0)^{-1}g_0(-x^\prime).
\end{equation}
Note that the $2\times2$ Nambu Green's function %$g(x,x^\prime)$ in Nambu space is denoted by $G_{B=0}(x,x^\prime)$ in Eq.~() of the main text.
\begin{equation}
g(x,x^\prime) \equiv G_{B=0}(x,x^\prime)
\end{equation}
where the latter notation is used in the main text.

For the purpose of ABS spectroscopy, we are interested in
$G(x_0,x_0,k_y,E+i0^+)$, obtained by setting $x=x^\prime=x_0$ in Eq.~\eqref{eq:supp:half_space_1}, where $x_0>0$ is a point near the boundary on the scale of the coherence length $\xi_0\sim v_F/\Delta$. The ABS appears as a subgap pole of this Green's function. Since $G_0(x)$ has no singularities for
$E<\min\!\left(|\Delta_+|,|\Delta_-|\right)$, any singularity must originate from
$G_0(0)^{-1}$ in the second term of Eq.~\eqref{eq:supp:half_space_1}. Thus, the existence of an ABS requires
\begin{equation}
    \text{det}\bigl[G_0(0)\bigr] = 0 \implies (E+i0^+)^2\bigl(\zeta^{-1}_+ + \zeta^{-1}_-\bigr)^2 = \bigl(\Delta_+\zeta^{-1}_+ + \Delta_-\zeta^{-1}_-\bigr)^2.
    \label{eq:supp:ABS_condition}
\end{equation}
As $E\rightarrow0^+$ ($E\rightarrow0^-$),
$\Delta_+\zeta_+^{-1}\rightarrow-i\,\mathrm{sgn}(\Delta_+)$
($+i\,\mathrm{sgn}(\Delta_+)$), and similarly for
$\Delta_-\zeta_-^{-1}$. Consequently, $E=0$ satisfies the above equation only if
\begin{equation}
    \text{sgn}\left(\Delta_+\right) = -\text{sgn}\left(\Delta_-\right).
    \label{eq:supp:ABS_criterion}
\end{equation}
Equation~\eqref{eq:supp:ABS_criterion} provides an explicit demonstration, in the extreme weak-coupling limit, of the criterion for the existence of zero-energy ABS. In the main text, we derive the same criterion in a more general setting and show that it is topologically protected.

Finally, we neglect
$\zeta_\pm/v_{Fx}\sim|\Delta_\pm|/v_{Fx}\sim1/\xi_0$
relative to
$k_{Fx}\sim k_F$
in the extreme weak-coupling limit, obtaining
\begin{equation}
    v_{Fx}G_0\left(x\right) = \Theta\left(x\right)\biggl(e^{ik_{Fx}x} \mathcal{P}_{+} + e^{-ik_{Fx}x} \mathcal{P}_{-}\biggr) + \Theta\left(-x\right)\biggl(e^{ik_{Fx}x} \mathcal{Q}_{+} + e^{-ik_{Fx}x} \mathcal{Q}_{-}\biggr)
    \label{eq:supp:G0(x)_2}
\end{equation}
\section*{IV. Zeeman Splitting of ABS}

At $B=0$, with appropriate basis choices for $S=0$ and $S=1$, the Hamiltonian is of the form $\mathcal{H}\left(\bm{k}\right) = \bm{u}(\bm{k})\cdot\bm{\tau}\otimes\sigma_0$, where $\bm{u}(\bm{k})\cdot\bm{\tau}$ is a $2\times 2$ matrix in Nambu ($\tau$) space. Consequently, the bulk GF $G_0(\bm{k},z) = g_0\left(\bm{k},z\right)\otimes \sigma_0$, and all derived objects such as $G_0(x,k_y,z) = g_0(x,k_y,z)\otimes\sigma_0$  and the half space GF, $G(x,x^\prime,k_y,z) = g(x,x^\prime,k_y,z)\otimes \sigma_0$ admit this factorization. Physically, the factorization implies a two-fold spin degeneracy of the quasiparticle and Andreev bound states. We investigate how the spin degeneracy is lifted for (1) spin singlet state and (2) spin triplet state.

\subsection*{1. Spin singlet}
For spin singlet SC in Zeeman field,
\begin{equation}
\mathcal{H}_B\left(\bm{k}\right) = \bm{u}(\bm{k})\cdot\bm{\tau}\otimes\sigma_0 + \tau_0\otimes\Vec{B}\cdot\Vec{\sigma}.
\label{eq:supp:HB_S=0}
\end{equation}
Defining $\mathbb{P}_\pm = \bigl(\sigma_0 \pm \hat{B}\cdot\Vec{\sigma}\bigr)/2$ and using $\sigma_0 = \mathbb{P}_+ + \mathbb{P}_-$ and $\hat{B}\cdot\Vec{\sigma} = \mathbb{P}_+ - \mathbb{P}_-$ the bulk GF,
\begin{eqnarray}
    G_0\left(\bm{k},z\right) &=& \bigl(z\mathbb{I}-\mathcal{H}_B(\bm{k})\bigr)^{-1}\nonumber\\
    &=& \biggl((z-B)\tau_0-\bm{u}(\bm{k})\cdot\bm{\tau}\biggr)^{-1}\otimes\mathbb{P}_+ + \biggl((z+B)\tau_0-\bm{u}(\bm{k})\cdot\bm{\tau}\biggr)^{-1}\otimes\mathbb{P}_-\nonumber\\
    &=& g_0\left(\bm{k},z-B\right)\otimes\mathbb{P}_+ + g_0\left(\bm{k},z+B\right)\otimes\mathbb{P}_-
    \label{eq:supp:G0(k)_singlet}
\end{eqnarray}
Fourier transforming with respect to $k_x$ the $\mathbb{P}_\pm$ decomposition of \eqref{eq:supp:G0(k)_singlet} is carried over to
\begin{equation}
    G_0(x,k_y,z) = g_0(x,k_y,z-B)\otimes\mathbb{P}_+ + g_0(x,k_y,z+B)\otimes\mathbb{P}_-
    \label{eq:supp:G0(x)_singlet}
\end{equation}
and finally using \eqref{eq:supp:half_space_1}, the half space GF
\begin{equation}
    G(x,x^\prime,k_y,z) = g(x,x^\prime,k_y,z-B)\otimes\mathbb{P}_+ + g(x,x^\prime,k_y,z+B)\otimes\mathbb{P}_-.
    \label{eq:supp:half_space_GF_singlet}
\end{equation}
Eqs.\eqref{eq:supp:G0(k)_singlet} and \eqref{eq:supp:half_space_GF_singlet} show that a Zeeman field applied to a singlet SC lifts the spin degeneracy by splitting all quasi-particle and ABS by $\pm B$, independent of the direction of $\Vec{B}$. Eq.\eqref{eq:supp:half_space_GF_singlet} is identical to eq.(9) in the main text given the identification of eq.(S34).
\subsection*{2. Spin triplet}
Recall that with $\Vec{d}(\bm{k}) = \Phi_1(\bm{k})\hat{d}$ and $\Vec{B}_\parallel = \bigl(\Vec{B}\cdot\hat{d}\bigr)\hat{d}$, $\Vec{B}_\perp = \Vec{B} - \bigl(\Vec{B}\cdot\hat{d}\bigr)\hat{d}$, the BdG hamiltonian for $S=1$ triplet SC in Zeeman field,
\begin{equation}
    \mathcal{H}_B(\bm{k}) = \bm{u}(\bm{k})\cdot\bm{\tau}\otimes \sigma_0 + \tau_0\otimes\Vec{B}_\parallel\cdot\Vec{\sigma} + \tau_z\otimes\Vec{B}_\perp\cdot\Vec{\sigma}.
    \label{eq:supp:HB_S=1}
\end{equation}
Bulk GF
\begin{eqnarray}
    G_0\left(\bm{k},z\right) &=& \bigl(z\mathbb{I}-\mathcal{H}_B(\bm{k})\bigr)^{-1}\nonumber\\
    &=& \frac{\Bigl(z\mathbb{I}+\mathcal{H}_B(\bm{k})\Bigr)\Bigl(\bigl(z^2-2\mathcal{S}(\bm{k})\bigr)\mathbb{I} + \mathcal{H}_B^2(\bm{k})\Bigr)}{\bigl(z^2-E^2_+(\bm{k})\bigr)\bigl(z^2-E^2_-(\bm{k})\bigr)}
    \label{eq:supp:G0(k)_triplet}
\end{eqnarray}
where $E_\pm(\bm{k})$ is defined in \eqref{eq:supp_spectrum} and $\mathcal{S}(\bm{k}) = \xi^2(\bm{k})+ \Phi_1^2(\bm{k})+B^2$.
To calculate $G_0(x)$ using \eqref{eq:supp:r-space-GF}, as before, we solve for the poles of $G_0(k_x)$ at fixed $k_y$ and $z=E+i0^+$. This is obtained by setting the denominator of \eqref{eq:supp:G0(k)_triplet} to zero which on some rearrangement gives
\begin{equation}
    \bigl(\xi^2(k_x) - \zeta^2_1(k_x)\bigr)\bigl(\xi^2(k_x) - \zeta^2_2(k_x)\bigr) = 0
    \label{eq:supp:pole_condition_S=1_a}
\end{equation}
where
\begin{equation}
    \zeta^2_{1,2}(k_x) = z^2 - \Phi^2_1(k_x) + B^2 \pm 2\sqrt{z^2B^2 - B^2_\perp\Phi^2_1(k_x)}
    \label{eq:supp:pole_condition_S=1_b}
\end{equation}
We solve eq.\eqref{eq:supp:pole_condition_S=1_a} in the extreme weak coupling regime $v_Fk_F/\Delta\gg1$. In this case we get eight poles: four of which are near $+k_{Fx}$ and four near $-k_{Fx}$ where $\xi(\pm k_{Fx}) =0$. We obtain this by expanding $\xi(k_x)$ and $\Phi_1(k_x)$ around $\pm k_{Fx}$ following \eqref{eq:supp:linear_dispersion}. The poles and the corresponding residues are tabulated in table~\ref{tab:residues2} and the symbols used are defined in the caption.
\begin{table}[t]
\centering
\renewcommand{\arraystretch}{1.8}
\begin{tabular}{c@{\hspace{2cm}}c}
\toprule
$k_0$ & $iv_{Fx}\,\mathrm{Res}\,G_0(k_0)$ \\
\midrule

\multicolumn{2}{c}{$\mathrm{Im}(k_0)>0$} \\
\midrule

$\displaystyle p_{1+} = +k_{Fx}+\frac{\zeta_{1+}}{v_{Fx}}$
&
$\displaystyle
\mathcal{P}_{1+}
=
i\frac{\mathcal{N}\!\left(\zeta_{1+},\Delta_+\right)}
{2\zeta_{1+}\bigl(\zeta^2_{1+} - \zeta^2_{2+}\bigr)}
$
\\[3ex]

$\displaystyle p_{2+} = +k_{Fx}+\frac{\zeta_{2+}}{v_{Fx}}$
&
$\displaystyle
\mathcal{P}_{2+}
=
i\frac{\mathcal{N}\!\left(\zeta_{2+},\Delta_+\right)}
{2\zeta_{2+}\bigl(\zeta^2_{2+} - \zeta^2_{1+}\bigr)}
$
\\[3ex]

$\displaystyle p_{1-} = -k_{Fx}+\frac{\zeta_{1-}}{v_{Fx}}$
&
$\displaystyle
\mathcal{P}_{1-}
=
i\frac{\mathcal{N}\!\left(-\zeta_{1-},\Delta_-\right)}
{2\zeta_{1-}\bigl(\zeta^2_{1-} - \zeta^2_{2-}\bigr)}
$
\\[3ex]

$\displaystyle p_{2-} = -k_{Fx}+\frac{\zeta_{2-}}{v_{Fx}}$
&
$\displaystyle
\mathcal{P}_{2-}
=
i\frac{\mathcal{N}\!\left(-\zeta_{2-},\Delta_-\right)}
{2\zeta_{2-}\bigl(\zeta^2_{2-} - \zeta^2_{1-}\bigr)}
$
\\[3ex]

\midrule

\multicolumn{2}{c}{$\mathrm{Im}(k_0)<0$} \\
\midrule

$\displaystyle q_{1+} = +k_{Fx}-\frac{\zeta_{1+}}{v_{Fx}}$
&
$\displaystyle
-\mathcal{Q}_{1+}
=
-i\frac{\mathcal{N}\!\left(-\zeta_{1+},\Delta_+\right)}
{2\zeta_{1+}\bigl(\zeta^2_{1+} - \zeta^2_{2+}\bigr)}
$
\\[3ex]

$\displaystyle q_{2+} = +k_{Fx}-\frac{\zeta_{2+}}{v_{Fx}}$
&
$\displaystyle
-\mathcal{Q}_{2+}
=
-i\frac{\mathcal{N}\!\left(-\zeta_{2+},\Delta_+\right)}
{2\zeta_{2+}\bigl(\zeta^2_{2+} - \zeta^2_{1+}\bigr)}
$
\\[3ex]

$\displaystyle q_{1-} = -k_{Fx}-\frac{\zeta_{1-}}{v_{Fx}}$
&
$\displaystyle
-\mathcal{Q}_{1-}
=
-i\frac{\mathcal{N}\!\left(\zeta_{1-},\Delta_-\right)}
{2\zeta_{1-}\bigl(\zeta^2_{1-} - \zeta^2_{2-}\bigr)}
$
\\[3ex]

$\displaystyle q_{2-} = -k_{Fx}-\frac{\zeta_{2-}}{v_{Fx}}$
&
$\displaystyle
-\mathcal{Q}_{2-}
=
-i\frac{\mathcal{N}\!\left(\zeta_{2-},\Delta_-\right)}
{2\zeta_{2-}\bigl(\zeta^2_{2-} - \zeta^2_{1-}\bigr)}
$
\\

\bottomrule
\end{tabular}
\caption{Poles of the $G_0(k_x)$ and the corresponding residues. For compactness we have defined $\zeta_{1\pm} = \Bigl((E+i0^+)^2-\Delta^2_\pm + B^2 + 2\sqrt{(E+i0^+)^2B^2-B^2_\perp\Delta^2_\pm}\,\Bigr)^{1/2}$and $\zeta_{2\pm} = \Bigl((E+i0^+)^2-\Delta^2_\pm + B^2 - 2\sqrt{(E+i0^+)^2B^2-B^2_\perp\Delta^2_\pm}\,\Bigr)^{1/2}$ with the branch of the square root chosen such that $\text{Im}\left(\zeta_{1\pm}\right), \text{Im}\left(\zeta_{2\pm}\right)>0$. The function in the numerator, $\mathcal{N}\left(\zeta,\Delta\right) = \biggl((E+i0^+)\mathbb{I} + \mathcal{H}\biggr)\biggl(\bigl((E+i0^+)^2-2\mathcal{S}\bigr)\mathbb{I} + \mathcal{H}^2\biggr)$, where $\mathcal{H}(\zeta,\Delta) = (\zeta\tau_z + \Delta \tau_x)\otimes \sigma_0 + \tau_0\otimes\Vec{B}_\parallel\cdot\Vec{\sigma} + \tau_z\otimes\Vec{B}_\perp\cdot\Vec{\sigma}$}
\label{tab:residues2}
\end{table}
Thus,
\begin{eqnarray}
    G_0(x) &=& \Theta(x)\biggl(\mathcal{P}_+^{(1)}~e^{ip^{(1)}_{+}x}+ \mathcal{P}_+^{(2)}~e^{ip^{(2)}_{+}x} + \mathcal{P}_-^{(1)}~e^{ip^{(1)}_{-}x} + \mathcal{P}_-^{(2)}~e^{ip^{(2)}_{-}x}\biggr) \nonumber\\
    && + \Theta(-x)\biggl(\mathcal{Q}_+^{(1)}~e^{iq^{(1)}_{+}x}+ \mathcal{Q}_+^{(2)}~e^{iq^{(2)}_{+}x} + \mathcal{Q}_-^{(1)}~e^{iq^{(1)}_{-}x} + \mathcal{Q}_-^{(2)}~e^{iq^{(2)}_{-}x}\biggr).
    \label{eq:supp:G0(x):S=1_a}
\end{eqnarray}
In the $k_F\xi_0 \gg 1$ limit, setting $p_+^{(1,2)},q_+^{(1,2)} \to +k_{Fx}$ and $p_-^{(1,2)},q_-^{(1,2)} \to -k_{Fx}$, eq.\eqref{eq:supp:G0(x):S=1_a} simplifies to
\begin{eqnarray}
    G_0(x) &=& \Theta(x)\Biggl(\biggl(\mathcal{P}_+^{(1)} + \mathcal{P}_+^{(2)}\biggr)~e^{ik_{Fx}x} + \biggl(\mathcal{P}_-^{(1)} + \mathcal{P}_-^{(2)}\biggr)~e^{-ik_{Fx}x}\Biggr) \nonumber\\
    && + \Theta(-x)\Biggl(\biggl(\mathcal{Q}_+^{(1)} + \mathcal{Q}_+^{(2)}\biggr)~e^{ik_{Fx}x} + \biggl(\mathcal{Q}_-^{(1)} + \mathcal{Q}_-^{(2)}\biggr)~e^{-ik_{Fx}x}\Biggr)
    \label{eq:supp:G00(x):S=1_b}
\end{eqnarray}
Using the \eqref{eq:supp:G00(x):S=1_b} and \eqref{eq:supp:half_space_1} we construct $G(x,x^\prime)$. For the special cases of $\Vec{B}\parallel\hat{d}$ and $\Vec{B}\perp\hat{d}$, the above equations can be used to show the decomposition of $G(x,x^\prime)$ in eqs. (9) and (10) of the main text.

\section*{V. SC order parameter in the original degrees of freedom}
In Eq.~1 of the main text, we express the pairing Hamiltonian in terms of the spin-degenerate IKS band operators $c_{\bm{k}\alpha}$ as
\begin{equation}
    \hat{H}_{\text{pair}}
    =
    \sum_{\bm{k}}
    \Delta_{\alpha\beta}(\bm{k})
    c_{\bm{k}\alpha}
    c_{-\bm{k}\beta}
    + \text{h.c.}
    \label{eq:supp:Hpair}
\end{equation}
Using Eq.~\eqref{eq:supp:IKS_band}, we rewrite the pairing Hamiltonian in terms of the original moir\'e-band degrees of freedom. This shows that pairing within a single IKS band inevitably generates finite-momentum Cooper pairs on both the graphene and the moir\'e scales. Substituting Eq.~\eqref{eq:supp:IKS_band} into Eq.~\eqref{eq:supp:Hpair} yields
\begin{align}
    \hat{H}_{\text{pair}}
    =\;& \sum_{\bm{k}} \Delta_{\alpha\beta}\!\left(\bm{k}\right)
        \biggl(\cos\!\Bigl(\tfrac{\theta_{\bm{k}}}{2}\Bigr)
            \psi_{\bm{k}+\bm{Q}/2,\,K,\alpha}
        + \sin\!\Bigl(\tfrac{\theta_{\bm{k}}}{2}\Bigr)
            \psi_{\bm{k}-\bm{Q}/2,\,K^\prime,\alpha}\biggr)
        \nonumber\\
    &\quad\times
        \biggl(\cos\!\Bigl(\tfrac{\theta_{-\bm{k}}}{2}\Bigr)
            \psi_{-\bm{k}+\bm{Q}/2,\,K,\beta}
        + \sin\!\Bigl(\tfrac{\theta_{-\bm{k}}}{2}\Bigr)
            \psi_{-\bm{k}-\bm{Q}/2,\,K^\prime,\beta}\biggr)
        + \text{h.c.}
    \label{eq:Hpair_2}\\
    =\;& \sum_{\bm{k}} \Delta_{\alpha\beta}\!\left(\bm{k}\right)
        \frac{1}{2}\sin\!\left(\theta_{\bm{k}}\right)
        \Bigl(
            \psi_{\bm{k}+\bm{Q}/2,\,K,\alpha}\,\psi_{-\bm{k}+\bm{Q}/2,\,K,\beta}
            +\psi_{\bm{k}-\bm{Q}/2,\,K^\prime,\alpha}\,\psi_{-\bm{k}-\bm{Q}/2,\,K^\prime,\beta}
        \Bigr)
    \nonumber\\
    &+ \sum_{\bm{k}} \Delta_{\alpha\beta}\!\left(\bm{k}\right)
        \Bigl(
            \cos^2\!\Bigl(\tfrac{\theta_{\bm{k}}}{2}\Bigr)
            \psi_{\bm{k}+\bm{Q}/2,\,K,\alpha}\,\psi_{-\bm{k}-\bm{Q}/2,\,K^\prime,\beta}
            +\sin^2\!\Bigl(\tfrac{\theta_{\bm{k}}}{2}\Bigr)
            \psi_{\bm{k}-\bm{Q}/2,\,K^\prime,\alpha}\,\psi_{-\bm{k}+\bm{Q}/2,\,K,\beta}
        \Bigr)
        + \text{h.c.}
    \label{eq:Hpair_3}
\end{align}
The second line of Eq.~\eqref{eq:Hpair_3} describes intervalley Cooper pairs with zero center-of-mass momentum. By contrast, the first line describes intravalley Cooper pairs with finite center-of-mass momentum $\bm{Q}$ on the moir\'e scale. The implications for the real-space modulation of the superconducting order parameter are discussed at the very end of the main text.

\end{document}